\documentclass[12pt]{iopart}
 
\expandafter\let\csname equation*\endcsname=\relax 
\expandafter\let\csname endequation*\endcsname=\relax 
\usepackage{amsmath,amssymb,amsthm} % note amssymb will load amsfonts per Barbara Beeton also to avoid physics use
\usepackage{braket}

\usepackage{euscript}
\usepackage{mathtools}  
\usepackage[dvipsnames]{xcolor}
\usepackage{graphicx}
\usepackage{csquotes}
\usepackage{anyfontsize}
\usepackage{dcolumn}
\usepackage{adjustbox}
\PassOptionsToPackage{hyphens}{url}
\usepackage{hyperref}
\usepackage{tikz}
\usepackage{hhline}
\usepackage{xcolor,colortbl}
\usepackage{xurl}
\usepackage{bm}

\definecolor{maroon}{cmyk}{0,0.87,0.68,0.32}
\DeclareMathAlphabet{\mathpzc}{OT1}{pzc}{m}{it}

\usepackage{array}
\usepackage{makecell}
\usepackage{caption}
\usepackage{subcaption}

\usepackage{bm}
\usepackage[inner = 30mm, outer = 20mm,  top = 30mm, bottom = 20mm, headheight = 13.6pt]{geometry}
\usepackage{tikz}
\usetikzlibrary{calc,fit,arrows.meta,calc,positioning}
\usepackage{xifthen}
\newcommand{\head}[1]{\textnormal{\textbf{#1}}}

\usepackage{color}
\usepackage[capitalize]{cleveref}

\begin{document}

\title{Thermodynamic Uncertainty Relations for Multipartite Processes}

\author{Gülce Kardeş}
\address{IPSP, University of Leipzig, Leipzig}
\ead{gulcekardes@gmail.com}
\author{David Wolpert}
\address{Santa Fe Institute, Santa Fe, New Mexico \\ Complexity Science Hub, Vienna \\
Arizona State University, Tempe, Arizona}
\ead{david.h.wolpert@gmail.com}

\vspace{10pt}
\begin{indented}
\item[]\date{\today}
\end{indented}
\begin{abstract}
The thermodynamic uncertainty relations (TURs) provide lower bounds on the entropy production (EP) of a system in terms of the statistical precision of an arbitrary current in that system. All conventional TURs derived so far have concerned a single physical system, differing from one another in what properties they require the system to have. However, many physical scenarios of interest involve multiple interacting systems, e.g. organelles within a biological cell. Here we show how to extend the conventional TURs to those scenarios.
A common feature of these extended versions of the TURs is that they bound the global EP, jointly generated by the set of interacting systems, in terms of a weighted sum of the precisions of the local currents generated within those systems -- plus an information-theoretic correction term. 
Importantly, these extended TURs can bound the global EP even when the global system does not 
meet any of the requirements of the conventional TURs.
After deriving these extended TURs we use them to obtain bounds that do not involve the global EP, but instead relate the local EPs of the individual systems and the statistical coupling among the currents generated
within those systems. We derive such bounds for both scalar-valued and vector-valued currents within each system. We illustrate our results with numerical experiments.
\end{abstract}

%\maketitle

\section{INTRODUCTION}
\label{sec:intro}

Stochastic thermodynamics has resulted in powerful results concerning the thermodynamic behavior of a broad spectrum of far-from-equilibrium systems, ranging from Brownian particles to molecular motors within living systems \cite{1}.
In particular, the thermodynamic uncertainty relations (TURs) are a set of results that 
provide upper bounds on the precision of a current generated by a dynamical system, in terms of the 
entropy production (EP) of that dynamics \cite{2,3,4,5,6,7,8,9,10,11,12}. 
There are many such TURs,
differing in their assumptions about the dynamics, e.g., whether it is a non-equilibrium steady state (NESS) \cite{2}, a time-homogeneous process \cite{6} or whether the driving protocol is time-symmetric \cite{5} or periodically time-dependent \cite{7}. It is also possible to derive bounds concerning the joint precision of multiple currents within a system \cite{8}.

While they vary from one another in many respects, all TURs derived so far have concerned a single physical system, coupled to its own set of one or more reservoirs. This strongly limits their applicability to modeling the physical world, since many physical scenarios of interest involve multiple interacting systems. In particular, in multipartite processes (MPPs) \cite{13, 14, 15} there are multiple distinct ``subsystems'' of an overarching system, each with its own set of reservoirs, independent of the reservoirs of the other subsystems. 

\begin{figure}
    \centering
     \begin{subfigure}{0.75\textwidth}
       \centering
       \includegraphics[scale=0.25]{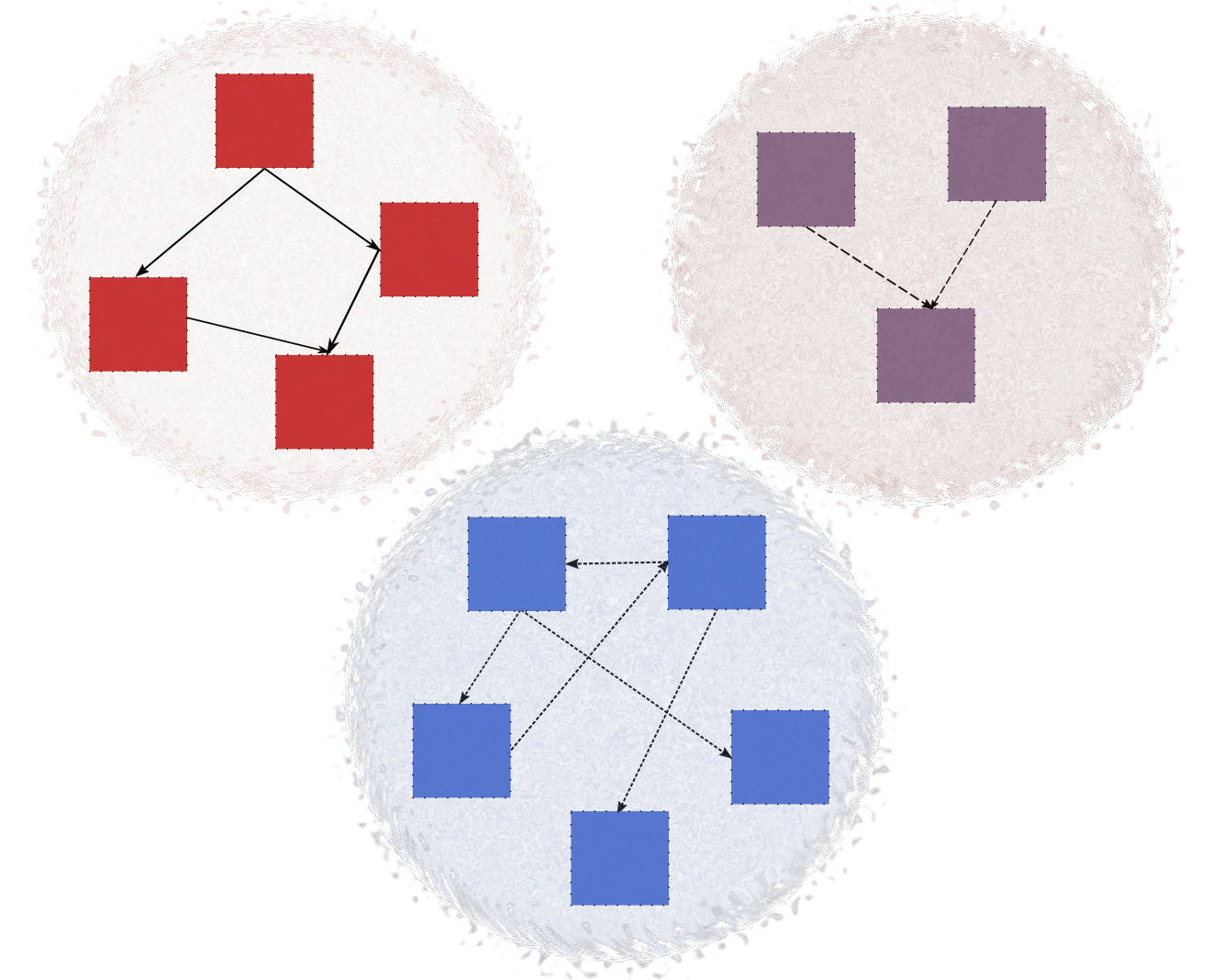}
       \caption{Protein complexes that are composed of subunits.}\label{fig:a}
    \end{subfigure}          
    \begin{subfigure}{0.64\textwidth}  
       \centering
       \includegraphics[scale=0.16]{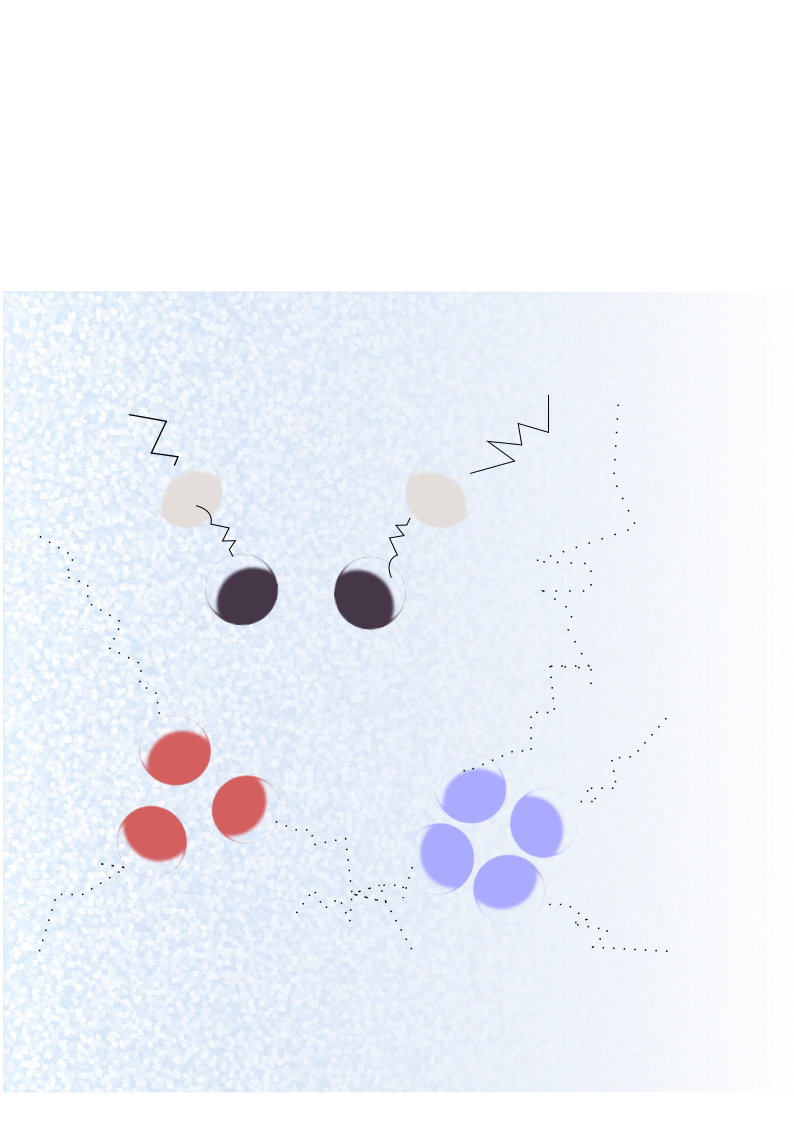}
       \caption{Cluster formation of self-propelling particles.}\label{fig:aa}
    \end{subfigure}    
    \caption{Examples of MPPs.}
\end{figure}

Fig.\,1 displays a set of scenarios that could be modeled as an MPP. 
First, note that the biological machinery inside a cell comprises multiple distinct
kinds of interactions between protein
complexes and molecules, suggesting a modular architecture \cite{16} whose dynamics
is governed by an MPP. For example, 
one might model a gene regulatory network by treating each of the protein complexes in
that network as a subsystem in an MPP. (In general, these subsystems could correspond to
protein complexes that regulate gene transcriptions 
and / or mediate the activity of enzymes.)
Such a network is illustrated figuratively in Fig.\,1 (a); see \cite{17} for a detailed analysis of the
information-thermodynamic dynamics of a tripartite example of such a regulatory network.

Fig.\,1 (b) illustrates another example of an MPP, involving cluster formation of self-propelling particles. As shown in \cite{18, 19} for the Kern-Frenkel model for Janus particles \cite{20}, cluster formation of self-propelling particles depend on the interaction range and a plenitude of thermodynamic parameters.  These dependencies can be captured in an MPP, where each Janus particle --- in the entire aggregate --- is a subsystem.

The goal of this paper is to extend the conventional TURs, applicable to single systems,
to settings of multiple, dependent systems. We begin in the next section by summarizing the conventional TURs that were derived earlier for single
systems~\cite{2, 5, 6, 8}. We continue with a review of rate matrix ``units", 
a key attribute of the thermodynamics of MPPs~\cite{21, 22, 23}. Intuitively, a unit is
a group of subsystems that collectively evolve in a way that is independent of all
other subsystems. 

Importantly, a particular subsystem can be in more than one unit in general. Such 
overlap among the units defines a graph, which captures the dependencies among the rate matrices
of the subsystems. This dependency graph plays a key role in the trajectory-level stochastic 
thermodynamics of MPPs. 
In particular, in~\cite{22} a trajectory-level decomposition of the global
EP of an MPP in terms of EPs of the individual units in the MPP is derived, 
where the form of the decomposition is determined by the dependency graph of the MPP. 
This decomposition is then used in~\cite{22} to derive two lower bounds on global EP, 
and these bounds are then used to extend the conventional TURs to MPPs.

In the following section we exploit the exact formula for expected EP from~\cite{22}, rather than
the lower bounds on EP that were investigated in~\cite{22}. 
This allows us to extend the conventional TURs in new ways.

These new extended TURs provide ``hybrid'' bounds on the {global} EP, in terms of the precisions of the
{local} currents generated within \textit{some} of the units, together with
the EPs of the remaining units --- plus an information-theoretic correction term.  This information-theoretic
correction term is a generalization of the multi-information among the states of the overall system,
reflecting the form of the MPP's dependency graph.

In the last section we illustrate our findings through numerical experiments involving multiple quantum dots. We conclude by discussing the implications of our results and suggesting possible future work. 
All proofs not in the main text are in the appendices.

\section{BACKGROUND}
\label{sec:background}

\subsection{Thermodynamic uncertainty relations}
\label{sec:previous_TURs_review}

A \textit{current} generated by any trajectory across a space $Y$ is any function of the form
\begin{align*}
J(\boldsymbol{x}) = \sum_{x \neq x^{\prime}} n_{x, x^{\prime}}(\boldsymbol{x})d(x, x^{\prime})
\tag{{\color{Black}{1}}} \label{1}
\end{align*}
where $n_{x, x^{\prime}}(\boldsymbol{x})$ corresponds to the total number of transitions from \textit{x} to \textit{y}, and $d(x, x^{\prime})=-d(x^{\prime}, x)$ is the anti-symmetric increment associated with a given transition.

Currents vary from one trajectory to another, and so the stochasticity of
the trajectories induces  a stochasticity of currents. This stochasticity 
can be quantified by the \textit{precision} of the current, i.e., by its variance to mean-square ratio. 
TURs provide upper bounds on the precision of \textit{any} current generated
by a system in terms of the associated expected EP. The TURs are independent of the precise choice
of $d$. In addition, which precise TUR applies to a given process only depends on
high-level properties of the driving protocol and of the 
associated distribution $\mathbf{P}$, independent of their details.

Historically, the first TUR \cite{2} took the form
\begin{align*}
\left<\sigma\right> \geq \frac{2{\left<J\right>}^{2}}{Var[J]}
\tag{{\color{Black}{2}}} \label{2}
\end{align*}
It applies whenever the system is in a NESS
throughout $[0, t_f]$. Note that this requires that the driving protocol be time-independent,
which limits the applicability of this bound. However, one can avoid
such a strong restriction on the physical scenario, at the expense of having a slightly weaker bound.
Assume that $\mathbf{P}$ evolves under a time-symmetric protocol, and that the starting distribution is identical to the ending distribution. In this case, the fluctuation theorem uncertainty relation (FTUR) \cite{5} holds
\begin{align*}
\left<\sigma\right>\geq \ln \left(\frac{2\left<J\right>^{2}}{Var[J]}+1\right) 
\tag{{\color{Black}{3}}} \label{3}
\end{align*}

More recently, a new type of TUR that applies to a combination
of \textit{instantaneous} currents and time-integrated currents was derived \cite{6}. This TUR holds independent of whether the distributions at the start and end of the interval of evolution are identical
\begin{align*}
\left<\sigma\right>\geq \frac{2\left<\tau
j^{\omega}_{t}\right>^{2}}{Var[J]}
\tag{{\color{Black}{4}}} \label{4}
\end{align*}

Note that these TURs consider single currents only, defined
by a single function $d(.,.)$. However, even
in a process that is not an MPP, in general there will be multiple currents,
defined by different functions $d(., .)$. Write  
$\boldsymbol{J}(\boldsymbol{x})$ for such a vector of currents, with expectation value
$\langle\boldsymbol{J}\rangle$. Also write $\boldsymbol{\Xi}^{-1}_{J}$ for the associated inverse covariance matrix. In \cite{8} 
the generalized Cramer-Rao inequality is used to derive a multi-dimensional 
version of the FTUR in terms of these quantitites, where $\boldsymbol{\Sigma}$ is an EP-like term:
\begin{align*}
\langle\boldsymbol{J}\rangle^{T} \boldsymbol{\Xi}_{J}^{-1}\langle\boldsymbol{J}\rangle \leq \frac{1}{2} \boldsymbol{\Sigma}
\tag{{\color{Black}{5}}} \label{5}
\end{align*}

Several papers have also investigated ways to unify many of the TURs and /
or extend them beyond CTMCs. For example,
\cite{4} considered the Hilbert space of any observables that are invariant under time-reversal. This let them derive a TUR-like bound that holds for all stationary Markov processes. 
As another example, \cite{10} showed how to
introduce the key variables of stochastic thermodynamics (such as the current and entropy production) into quantum field theory, and then expressed the TUR given by \cite{2}
in field-theoretic terms. 

The discovery of TURs inspired the derivation of many other bounds that constrain the precision of other observables. In particular, \cite{24} showed by building on \cite{25} that the distribution for the time to first hit a large threshold current must satisfy a TUR-like relation.

\subsection{Rate matrix units}
\label{sec:rate_matrix_units}

\begin{figure}[b]
\centering
     \begin{subfigure}{0.44\textwidth}
       \centering
       \includegraphics[scale=0.4]{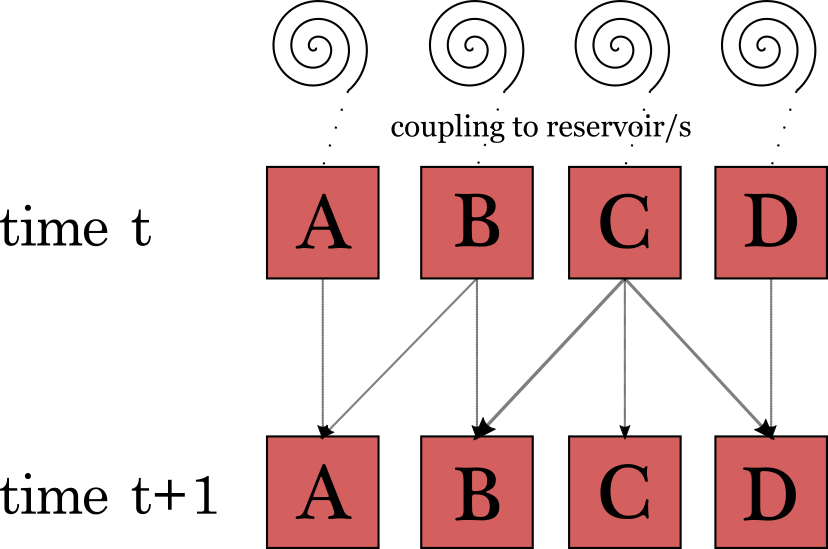}
       \caption{Four subsystems that are coupled to their own sets of one or more reservoirs interact in an MPP. The arrows dictate the dependencies in the associated rate matrices.}\label{fig:a}
    \end{subfigure}          
    \quad % spacing between the subfigures
    \begin{subfigure}{0.44\textwidth}  
       \centering
       \includegraphics[scale=0.4]{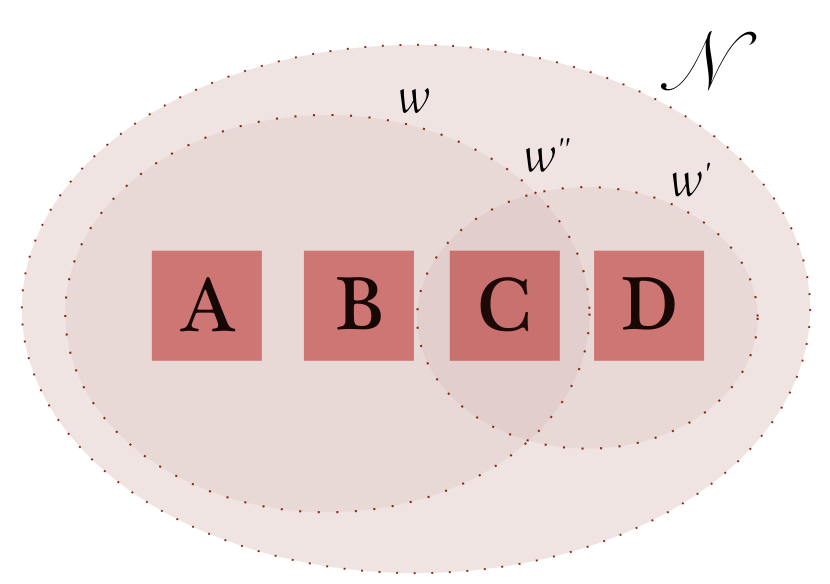}
       \caption{The three overlapping sets specify the three \textit{units} of the MPP illustrated above.}\label{fig:aa}
    \end{subfigure}    
    \caption{A representative MPP.}
\end{figure}

We write $\mathcal{N}$ for a set of $N$ subsystems with finite state spaces $\left\{X_{i}: i=1, \ldots N\right\}$. $X$ is the joint space of $\mathcal{N}$, and $x$ is a vector in $X$. For any $A \subset \mathcal{N}$, we write $-A:=\mathcal{N} \backslash A$ for the complement. The full distribution over 
$X$ at time $t$ is denoted by $p^{X}(t)$, with the particular value for state $x$ written
as $p_x(t)$. We use $\delta(., .)$ to indicate the Kronecker delta function.

We write $\textit{\textbf{X}}$ for the set of all possible trajectories of the system in $\left[0, t_{f}\right]$, a time interval of interest where the system evolves. A particular trajectory of states is written as $\textit{\textbf{x}}$, and the probability density for such a trajectory $\textit{\textbf{x}}$ is written as $\mathbf{P}(\boldsymbol{x})$. In general,
the Shannon entropy of a distribution is written as $S(.)$, e.g., $S\left(p^{X}(t)\right)$
is the Shannon entropy of the distribution over joint states at time $t$.

In an MPP, the global system's evolution is determined by a continuous-time Markov chain (CTMC)
given by a set of time-varying stochastic rate matrices, $\left\{W_{x}^{x^{\prime}}(i ; t): i=1, \ldots, N\right\}$, where for all $i$, $W_{x}^{x^{\prime}}(i ; t)=0$ if
$x_{-i}^{\prime} \neq x_{-i}$ \cite{13, 21, 22, 23}:
\begin{align*}
\frac{d p_{x}(t)}{d t} &=\sum_{x^{\prime}} W_{x}^{x^{\prime}}(t) p_{x^{\prime}}(t)
\tag{{\color{Black}{6}}} \label{6}
\\
&=\sum_{x^{\prime}} \sum_{i \in \mathcal{N}} W_{x}^{x^{\prime}}(i ; t) p_{x^{\prime}}(t)
\tag{{\color{Black}{7}}} \label{7}
\end{align*}
We refer to the (pre-fixed) trajectory of the matrices $W(i;t)$ across the interval
$[0, t_f]$ as the \textit{driving protocol} of the process. For any $A \subseteq \mathcal{N}$, we define
\begin{align*}
W_{x}^{x^{\prime}}(A ; t):=\sum_{i \in A} W_{x}^{x^{\prime}}(i ; t)
\tag{{\color{Black}{8}}} \label{8}
\end{align*}

For each subsystem $i$, we write $r(i;t)$ for any set of subsystems at time $t$ that includes $i$, such that it is possible to write
\begin{align*}
W_{x}^{x^{\prime}}(i ; t)=W_{x_{r(i ; t)}}^{x_{r(i ; t)}^{\prime}}(i ; t) \delta\left(x_{-r(i ; t)}^{\prime}, x_{-r(i ; t)}\right)
\tag{{\color{Black}{9}}} \label{9}
\end{align*}
for appropriate functions $W_{x_{r(i ; t)}}^{x_{r(i ; t)}^{\prime}}(i;t)$.
$r(i;t)$ will not be unique in general. 

We use the term \textit{unit} to refer to any set of subsystems
$\omega$ such that $i \in \omega$ implies $r(i;t) \subseteq \omega$. 
Note that any intersection of units is a unit, and any union of units is a unit. 
A set of units that covers $\mathcal{N}$ and that is closed under intersections is referred to
as a \textit{unit structure}, denoted by $\mathcal{N}^{*}$.
In the following we consider only the unit structures whose topology does not change in time, and we require that $\mathcal{N}$ itself is not a part of the unit structure. Fig.\,2 depicts a representative form of such an MPP that is structured by \cref{1,2,3,4}.
Note that the most fine-grained unit structure is given by the transitive closure
of the $r(.)$ relation. For example, in Fig.\,2, the unit that contains subsystem $A$
is $\{A, B, C\}$, since $B \in r(A)$ and $C \in r(B)$.

\subsection{Trajectory-level thermodynamics of MPPs}
\label{sec:trajectory.level_thermodynamics_mpp}

\colorlet{A}{green!10!red!10!blue!10!}
\colorlet{B}{green!10!red!10!blue!10!}
\colorlet{C}{green!10!red!10!blue!10!}

\pgfdeclarelayer{background}
\pgfdeclarelayer{foreground}
\pgfsetlayers{background,main,foreground}

\tikzset{
timeline/.style={arrows={}}%
,timeline style/.style={timeline/.append style={#1}}%
,year label/.style={font=\small\bfseries,below}%                  <- removed \sffamily
,year label style/.style={year label/.append style={#1}}%
,year tick/.style={tick size=0pt}%
,year tick style/.style={year tick/.append style={#1}}%
,minor tick/.style={tick size=0pt, very thin}%
,minor tick style/.style={minor tick/.append style={#1}}%
,period/.style={solid,line width=\timelinewidth,line cap=square}%
,periodbox/.style={font=\small\bfseries,text=black}%              <- removed \sffamily
,eventline/.style={draw,red,thick,line cap=round,line join=round}%
,eventbox/.style={rectangle,rounded corners=3pt,inner sep=3pt,fill=red!25!white,text width=3cm,anchor=west,text=black,align=left,font=\small}%
,tick size/.code={\def\ticksize{#1}}%
,labeled years step/.code={\def\yearlabelstep{#1}}%
,minor tick step/.code={\def\minortickstep{#1}}%
,year tick step/.code={\def\yeartickstep{#1}}%
,enlarge timeline/.code={\def\enlarge{#1}}%
,eventboxa/.style={eventbox,text width=#1,draw=A,fill=none}%
,eventboxb/.style={eventbox,text width=#1,draw=A,fill=none}%
}
\newcommand*{\drawtimeline}[5][]{%
\def\fromyear{#2}%
\def\toyear{#3}%
\def\timelinesize{#4}%
\def\timelinewidth{#5}%
\pgfmathsetmacro{\timelinesizept}{\timelinesize}%
\pgfmathsetmacro{\timelinewidthpt}{\timelinewidth}%
\pgfmathsetmacro{\timelineoffset}{\timelinewidth/2}%
\pgfmathsetmacro{\timelineoffsetpt}{\timelineoffset}%

\begin{scope}[x=1pt, y=1pt, 
    labeled years step=1,
    minor tick step=0.25,
    enlarge timeline=0cm,
    year tick step=1,#1]
    \pgfmathsetmacro{\enlargept}{\enlarge}
    \pgfmathsetmacro{\yearticksep}{\timelinesize/((\toyear-\fromyear)/\yeartickstep)}
    \pgfmathsetmacro{\minorticksep}{\timelinesize/((\toyear-\fromyear)/\minortickstep)}
    \pgfmathsetmacro{\minorticklast}{\minorticksep/\minortickstep}
    \foreach \y[remember=\y as \lasty (initially 0), count=\i from \fromyear] in {0,\yearticksep,...,\timelinesizept}{
        \coordinate (Y-\i) at (\y,0);
        \draw[year tick] (\y,-\ticksize/2) --  ++(0,\ticksize);
        \ifnum\i=\toyear\breakforeach\else
        \foreach \q[count=\j from 0] in {0,\minorticksep,...,\minorticklast}
{
            \coordinate (Y-\i-\j) at (\q+\y,0);
                \draw[minor tick] (\q+\y,-\ticksize/2) -- ++(0,\ticksize);
        };\fi};%
    \pgfmathsetmacro{\nextyear}{int(\fromyear+\yearlabelstep)}
        \draw[timeline] (0,0) -- ++(-\enlargept,0) (0,0) -- ++(\timelinesizept,0) coordinate (end) -- ++(\enlargept,0);
\end{scope}

}

\newcommand{\period}[5]{\draw[period,#1] (Y-#2) -- (Y-#3) node[periodbox,#5,midway,text=white] {#4};}

\newcommand{\vevent}[9]{
     \pgfmathtruncatemacro{\syr}{#2}
     \pgfmathtruncatemacro{\smth}{#3-1}
     \pgfmathsetmacro{\dim}{#4/31}
     \ifthenelse{#3=12}{%
        \pgfmathtruncatemacro{\fyr}{#2+1}
        \pgfmathtruncatemacro{\fmth}{0}
        }{%
        \pgfmathtruncatemacro{\fyr}{#2}
        \pgfmathtruncatemacro{\fmth}{#3}
        }
     \draw[eventline,#1]($(Y-\syr-\smth)!\dim!(Y-\fyr-\fmth)$) -- ++(#5) -- ++(#6) node[#7] (#8) {#9};
     }
\tikzset{
    block/.style 2 args = {
        draw=none, inner sep=0, outer sep=0,
        rounded corners=3pt,
        fit=(#1) (#2)}
}

\newcommand{\fnode}[4][]{
    \coordinate (bottom left) at (#2);
    \coordinate (top right) at (#3);
     \node[block={bottom left}{top right}, #1, label=center:#4] {};
}
\begin{figure*}[t]
\centering
\resizebox{16cm}{7.2cm}{%
\begin{tikzpicture}
\drawtimeline[
labeled years step=1,
minor tick step=0.083333,
timeline style={draw=green!10!red!10!blue!10!,line width=\timelinewidthpt},
minor tick style={-,green!10!red!10!blue!10!,tick size=0pt,line width=0pt,yshift=-\timelineoffsetpt},
]%
{2017}{2019}{19cm}{0.5cm};
\period{A}{2017-0}{2017-1}{\color{Black}2015}{}
\period{A}{2018-6}{2018-7}{\color{Black}2020}{}
\period{A}{2018-7}{2018-8}{}{}
\begin{pgfonlayer}{background}
\vevent{A}{2017}{1}{10}{90:1.2cm}{45:1.2cm}{eventboxa=5cm,anchor=west}{H}{Steady-state TUR, applicable to NESS, though excludes many classes of driving protocols \cite{2}. \begin{center}\color{green!10!blue!80!red!60!}
$\frac{\operatorname{Var}[J]}{\langle J\rangle^{2}} \geq \frac{2}{\langle\sigma\rangle}$
\end{center}}
\vevent{A}{2017}{6}{8}{-90:2cm}{45:-0.5cm}{eventboxa=3.3cm,anchor=east}{H}{FTUR from (joint) fluctuation theorem of EP and currents \cite{5}. \begin{center}\color{green!10!blue!80!red!60!}$\frac{\operatorname{Var}[J]}{\langle J\rangle^{2}} \geq \frac{2}{e^{\langle\sigma\rangle}-1}$\end{center}
}
\vevent{A}{2017}{12}{18}{90:3.0cm}{45:0.4cm}{eventboxa=3cm,anchor=west}{H}{Multidimensional TURs \cite{8}. \begin{center}\color{green!10!blue!80!red!60!}$\langle\boldsymbol{J}\rangle^{T} \boldsymbol{\Xi}_{J}^{-1}\langle\boldsymbol{J}\rangle \leq \frac{1}{2} \boldsymbol{\Sigma}$\end{center}}
\vevent{A}{2017}{12}{5}{-90:2cm}{45:-1.25cm}{eventboxa=3.4cm,anchor=east}{H}{A set of other trade-off relations concerning different observables, including the first passage times \cite{24, 25}.
\begin{center}\color{green!10!blue!80!red!60!}
$\frac{\operatorname{Var}(T)}{\langle T\rangle^{2}} \geq \frac{2}{\langle\sigma\rangle}$
\end{center}}
\vevent{A}{2018}{2}{1}{-90:2.5cm}{45:0.4cm}{eventboxa=3cm,anchor=west}{H}{TURs for arbitrary initial states (in time-homogeneous processes) \cite{6}.
\begin{center}\color{green!10!blue!80!red!60!}$\frac{\operatorname{Var}[J]}{(T j(T))^{2}} \geq \frac{2}{\langle\sigma\rangle}$\end{center}}
\vevent{A}{2018}{6}{14}{90:2cm}{45:0.5cm}{eventboxa=3cm,anchor=west}{H}{TURs constructed in Hilbert space \cite{4};  Field-theoretic TURs \cite{10}.}
\end{pgfonlayer}
\end{tikzpicture}}
\caption{A brief history of TURs.}
\label{fig:timeline}
\end{figure*}
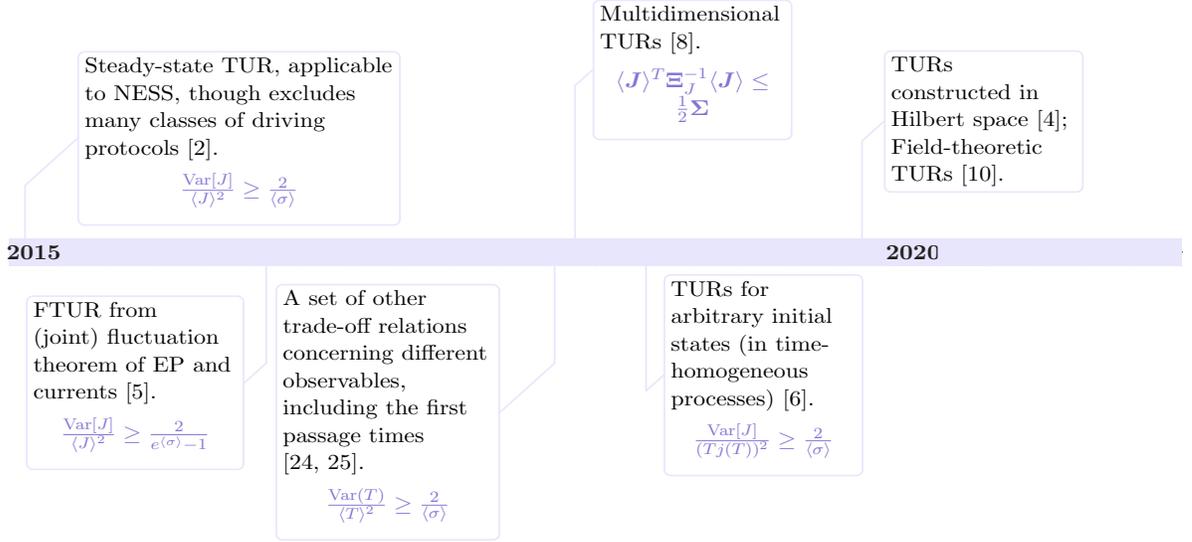

Physically, in general, in an MPP each subsystem is coupled to its own set of one or more reservoirs, and no two reservoirs can interact \cite{13, 21, 22, 23}. In our analysis of the MPP TURs, we allow the full generality of either one reservoir per subsystem
or multiple reservoirs per subsystem, except in the special case of an NESS, where the associated subsystems must have multiple reservoirs.

Choosing units so that $k_{B}=1$, and assuming LDB,
the (local) stochastic entropy of any set of subsystems $\alpha$ is \cite{13, 21, 22, 23}:
\begin{align*}
s^{\alpha}\left(x_{\alpha}(t)\right):=-\ln p_{\boldsymbol{x}_{a}(t)}(t)
\tag{{\color{Black}{10}}} \label{10}
\end{align*}
with its change from $t=0$ to $t=t_{f}$ written as
\begin{align*}
\Delta s^{\alpha}\left(x_{\alpha}\right):=\ln p_{x_{a}(0)}(0)-\ln p_{x_{a}\left(t_{f}\right)}\left(t_{f}\right)
\tag{{\color{Black}{11}}} \label{11}
\end{align*}

Let $\alpha$ be a set of subsystems (not necessarily a unit).
We use $Q^\alpha(\boldsymbol{x})$ to indicate the \textit{local} entropy 
flow (EF) into the subsystems in $\alpha$ from their
reservoirs during the interval $[0, t_f]$ if the full system follows trajectory $\boldsymbol{x}$. 
(See Appendix \,A for formal details.) The \textit{local} EP of any unit $\omega$ is just the difference between the
associated local EF and change in stochastic entropy:
\begin{align*}
\sigma^{\omega}(\boldsymbol{x}):=\Delta s^{\omega}(\boldsymbol{x})-Q^{\omega}(\boldsymbol{x})
\tag{{\color{Black}{12}}} \label{12}
\end{align*}
The \textit{global} EP is the special case where $\omega = \mathcal{N}$:
\begin{align*}
\sigma(\boldsymbol{x}):=\Delta s(\boldsymbol{x})-Q(\boldsymbol{x})
\tag{{\color{Black}{13}}} \label{13}
\end{align*}
From now on we leave the specification $\mathcal{N}$ implicit if it is clear from context. 

Given any unit-indexed function $f^{\omega}: X \rightarrow \mathbf{R}$,
the associated \textit{inclusion-exclusion sum} is defined as
\begin{align*}
\widehat{\sum_{\omega^{\prime} \in \mathcal{N}}} f^{\omega^{\prime}}(\boldsymbol{x}):=&\sum_{j=1}^{n} f^{\omega_{j}}(\boldsymbol{x})-\sum_{1 \leq j<j^{\prime} \leq n} f^{\omega_{j} \cap \omega_{j^{\prime}}(\boldsymbol{x})} \\
&+ \sum_{1 \leq j<j^{\prime}<j^{\prime \prime} \leq n} f^{\omega_{j} \cap \omega_{j}^{\prime} n \omega_{j^{\prime}}}(\boldsymbol{x})-\ldots
\tag{{\color{Black}{14}}} \label{14}
\end{align*}
Building on this we define the \textit{in-ex information} over units during $[0, t_f]$ as
\begin{align*} \mathcal{I}^{\mathcal{N}^{*}}(\boldsymbol{x}(t)):=&\left[\widehat{\sum_{\omega \in \mathcal{N}^{*}}} s^{\omega}(\boldsymbol{x}(t))\right]-s(\boldsymbol{x}(t))
\tag{{\color{Black}{15}}} \label{15}
\\ =&-s(\boldsymbol{x}(t))+\sum_{j=1}^{n} s^{\omega_{j}}(\boldsymbol{x}(t)) \\
& -\sum_{1 \leq j<j^{\prime} \leq n} s^{\omega_{j} \cap \omega_{j^{\prime}}}(x(t))+\ldots 
\end{align*}
Note that the in-ex information is equivalent to multi-information in the case of independent units. In particular, if the unit structure consists solely of two units that are independent, then \cref{10} simply corresponds to the mutual information between these units.

It is shown in \cite{21} that the global EP is given by the in-ex sum of the local EPs minus the change in in-ex information:
\begin{align*}
\sigma(\boldsymbol{x})=\widehat{\sum_{\omega \in \mathcal{N}^{*}}} \sigma^{\omega}(\boldsymbol{x})-\Delta \mathcal{I}^{\mathcal{N}^{*}}(\boldsymbol{x})
\tag{{\color{Black}{16}}} \label{16}
\end{align*}
\cref{11} is a keystone of our analysis below.

\section{MAIN RESULTS}
\label{sec:main_results}

As mentioned in the introduction,
one major limitation of the TURs reviewed above
is that they only apply to single systems, evolving
in isolation, and so do not apply to MPPs in particular. To extend them to forms
that apply to MPPs, first we define currents over units in a way that parallels \cref{12}:
\begin{align*}
J^{\omega}(\boldsymbol{x}) &= J^{\omega}(\boldsymbol{x}_\omega) \\
    &= \sum_{x_\omega \neq x^{\prime}_\omega} n^{\omega}_{x_\omega, x^{\prime}_\omega}(\boldsymbol{x}_\omega)d_\omega(x_\omega, x^{\prime}_\omega)
\tag{{\color{Black}{17}}} \label{17}
\end{align*}
where $d_\omega(., .)$ is an arbitrary anti-symmetric function. Importantly, because each unit evolves autonomously, as a self-contained
CTMC, each local EP $\sigma^{\omega}$ bounds the current $J^{\omega}$ generated in the associated unit $\omega$ according to the previously derived TURs of single systems. Building on this observation, the central mathematical tool that we use here is given by averaging both sides of \cref{11} according to ${\textbf{P}_{\textit{\textbf{x}}_\omega}}$: 
\begin{align*}
\left<\sigma\right>={\widehat{\sum_{\omega \in \EuScript{N^*}}}\left<\sigma^{\omega}\right>- \left<\Delta \EuScript{I}^{N^*}\right>_{\textbf{P}_{\textit{\textbf{x}}_\omega}}}
\tag{{\color{Black}{18}}} \label{18}
\end{align*}
\noindent
The first term on the RHS of the above equation is an in-ex sum of the expected local EPs over individual units, each of which evolves according to its own self-contained CTMC. As the decomposition in \cref{18} is exact, it is plausible to extend the TURs to get a range of allowable values of global EP in terms of unit currents and the change in in-ex information. Note that the (in-ex sum of the) local EPs over units can be bounded by the precision of the (in-ex sum of the) local current precisions generated in the corresponding units, given that the resulting range of global EP values are non-negative. To analyze this, we portion the in-ex sum into two standard sums: one which might potentially contribute with negative terms, and the remaining would be strictly composed of positive terms. We capture this by \cref{19}, where ${\sum}_{\omega \in \EuScript{N}_{o}^{*}}$ and ${\sum}_{\omega \in \EuScript{N}_{e}^{*}}$ denotes the positive and negative contributions to the in-ex sum, respectively. The positivity (negativity) of the contributions are assigned by the odd (even) parity of the intersecting units that form each element of the in-ex sum on the RHS of \cref{18}.
\begin{align*}
\left<\sigma\right>={\sum_{\omega \in \EuScript{N}_{o}^{*}}}\left\langle\sigma^{\omega}\right\rangle-{\sum_{\omega \in N_{e}^{*}}}\left\langle\sigma^{\omega}\right\rangle-\widehat{\sum_{\omega \in \EuScript{N}^{*}}} \left<\Delta \EuScript{I}^{\EuScript{N}^{*}_{e}}\right>_{\textbf{P}_{\textit{\textbf{x}}_\omega}}
\tag{{\color{Black}{19}}} \label{19}
\end{align*}

Our procedure for extending the TURs to apply to MPPs strongly depends on this property.
\begin{table*}[h!]
\begin{adjustbox}{width=1.00\textwidth}
{\normalsize{\begin{tabular}{>{\centering\arraybackslash}p{20cm}}
    \rowcolor{White}
    \head{\color{Black}\Large{\textrm{TURs \& MPP TURs}}} \\ 
    %\multicolumn{2}{c}{\cellcolor{black!60}\head{TEMPERATURE SENSOR}} \\
    \hline \thead{{\large{\color{blue!80!}Steady-state TUR} \cite{2}} \large{i.e..,  system is in a NESS as it evolves:} \\ \Large{${\langle\sigma\rangle} \geq \frac{2\langle J\rangle^{2}}{\operatorname{Var}(J)}$} \\ {\large{\color{red!80!}MPP Steady-state TURs}} \large{i.e..,  each unit is in a NESS as it evolves:} \\ \Large{{$\qquad \: {\langle\sigma\rangle} \geq \sum_{\omega \in \EuScript{N}_{o}^{*}} \frac{2\left\langle J^{\omega}\right\rangle^{2}}{\operatorname{Var}\left[J^{\omega}\right]}-\sum_{\omega \in \EuScript{N}_{e}^{*}}\left\langle\sigma^{\omega}\right\rangle$} \,\,\,\,\,\qquad\qquad\,\quad \cref{21}} \\ \Large{${\left\langle\sigma \mid \sigma^{\omega}\right\rangle} \geq \sum_{\omega' \in \omega^{*}_{o}} \frac{2\left<J^{\omega'}\right>^{2}}{Var[J^{\omega'}]} - \sum_{\omega^{\prime}\in {\omega}_{e}^{*}}\left\langle\sigma^{\omega}\right\rangle $\qquad \qquad \qquad \cref{35}}} \\ 
    \hline \thead{{\large{{\color{blue!80!}FTUR} \cite{5} \large{i.e..,  $p_{0}$ and $p_{f}$ are identical}, \large{$\lambda$ is time-symmetric:}}} \\ \Large{$\qquad {\left<\sigma\right>}\geq \ln \left(\frac{2\left<J\right>^{2}}{Var[J]}+1\right)$} \\ \large{{\color{red!80!}MPP FTURs}} \large{i.e..,  for each unit, $p_{0}^{\omega}$ and $p_{f}^{\omega}$ are identical, $\lambda^{\omega}$ is  time-symmetric:}} \\ \Large{\quad\, $\left<\sigma\right>\geq \sum_{\omega \in \EuScript{N}^{*}_{o}} \ln \left(\frac{2\left<J^{\omega}\right>^{2}}{Var[J^{\omega}]} + 1\right) -\sum_{\omega \in \EuScript{N}_{e}^{*}}\left\langle\sigma^{\omega}\right\rangle$ \quad \, \ \cref{24}} \\ \Large{\,$\left\langle\sigma \mid \sigma^{\omega}\right\rangle\geq \sum_{\omega' \in \omega^{*}_{o}} \ln \left(\frac{2\left<J^{\omega'}\right>^{2}}{Var[J^{\omega'}]} + 1\right) - \sum_{\omega^{\prime}\in {\omega}_{e}^{*}}\left\langle\sigma^{\omega}\right\rangle$ \ \cref{36}}  \\ 
    \hline \thead{{\large{\color{blue!80!}TUR with instantaneous current} \cite{6}} \large{i.e..,  system evolves} \large{time-homogeneously:} \\ \Large{${\langle\sigma\rangle} \geq \frac{2 \tau\left\langle j_{\tau}\right\rangle^{2}}{\operatorname{Var}(J)} $} \\ {\large{\color{red!80!}MPP TURs with instantaneous current}} \large{i.e..,  each unit evolves time-homogeneousy:}} \Large{$\qquad \ {\left<\sigma\right>}\geq \sum_{\omega \in \EuScript{N}^{*}_{o}}\frac{2\left<\tau^{\omega}
j^{\omega}_{t}\right>^{2}}{Var[J^{\omega}]} - \sum_{\omega \in \EuScript{N}_{e}^{*}}\left\langle\sigma^{\omega}\right\rangle$ \quad \ \qquad \, \,\cref{27}} \\ \Large{\,${\left\langle\sigma \mid \sigma^{\omega}\right\rangle} \geq \sum_{\omega' \in \omega^{*}_{o}}\frac{2\left<\tau^{\omega'}
j^{\omega'}_{t}\right>^{2}}{Var[J^{\omega'}]} - \sum_{\omega^{\prime}\in {\omega}_{e}^{*}}\left\langle\sigma^{\omega}\right\rangle $ \qquad\quad\cref{37}} \\
    \hline \thead{\large{{{\color{blue!80!}Steady-state MTUR}} \cite{8}} \large{i.e.., multi-dimensional system is in a NESS as it evolves:} \\ \Large{$\langle \sigma \rangle \geq 2\langle\boldsymbol{J}\rangle^{T} \Xi_{J}^{-1}(\boldsymbol{J}\rangle$} \\
     \large{{\color{red!80!}MPP Steady-state MTURs}} \large{i.e..,  each 
     multi-dimensional unit is in a NESS as it evolves:} 
     \\  \Large{${\sum}_{w} \sigma^{\omega}(\textit{\textbf{x}}_\omega)-\Delta \mathcal{I}^{\EuScript{N^*}}(\textit{\textbf{x})} \geq 2\langle\boldsymbol{J}\rangle^{T} \Xi_{J}^{-1}\langle\boldsymbol{J}\rangle$ \quad \quad \, \,\,\,\,\,\, \, \cref{44}} \\ \Large{$\quad \, D\left(\mathbf{P}\left(\vec{\sigma}^{\mathcal{A}}\right) \| \tilde{\mathbf{P}}\left(-\vec{\sigma}^{\mathcal{A}}\right)\right) \geq  2\langle\boldsymbol{J}^{\mathcal{A}}\rangle^{T}{\Xi_{J^{\mathcal{A}}}^{-1}}\langle\boldsymbol{J}^{\mathcal{A}}\rangle$ \qquad \cref{50}}} \\
    \hline \thead{\large{{{\color{blue!80!}MFTUR}} \cite{8}} 
    \large{i.e.,  $p_{0}$ and $p_{f}$ are identical,} \large{$\lambda$ is  time-symmetric:} \\ \Large{$\Sigma \geq 2\langle\boldsymbol{J}\rangle^{T} \Xi_{J}^{-1}(\boldsymbol{J}\rangle$} \\ \large{{\color{red!80!}MPP MFTURs}} \large{i.e.,  for each unit, $p_{0}^{\omega}$ and $p_{f}^{\omega}$ are identical, $\lambda^{\omega}$ is  time-symmetric:} \\
    \Large{\qquad \qquad \qquad \quad $\frac{\operatorname{Covar}\left[J^{\alpha_{i}}, J^{\beta_{j}}\right]}{\left\langle J^{\alpha_{i}}\right\rangle\left\langle J^{\beta_{j}}\right\rangle} \geq \frac{2 \cdot \left(1+e^{\langle\sigma^{\cup \mathcal{A}}-\sigma^{\cup \mathcal{\mathcal{B}}}\rangle}\right)}{e^{\langle\sigma^{\cup \mathcal{\mathcal{A}}}\rangle}+e^{\langle\sigma^{\cup \mathcal{\mathcal{B}}}\rangle}-\left(1+e^{\langle\sigma^{\cup \mathcal{\mathcal{A}}}-\sigma^{\cup \mathcal{\mathcal{B}}}\rangle}\right)}$ 
            \cref{41}} \\ 
    \Large{\,\qquad \qquad \qquad \qquad \qquad \ $\frac{\operatorname{Var} J^{\alpha_{i}}}{\left\langle J^{\alpha_{i}}\right\rangle^{2}}  \geq \frac{2}{e^{\langle\sigma^{\cup \mathcal{A}}\rangle}-1}$ \qquad \qquad \qquad \qquad \qquad \ \cref{42}}
    \\ \Large{\qquad \qquad \, \,\qquad \quad \qquad \ $\langle\sigma^{\cup \mathcal{A}}\rangle \geq  \ln\left(2\langle\boldsymbol{J}^{\mathcal{A}}\rangle^{T}{\Xi_{J^{\mathcal{A}}}^{-1}}\langle\boldsymbol{J}^{\mathcal{A}}\rangle+1\right)$ \, \cref{46}} \\\Large{\ \ $D\left(\mathbf{P}\left(\vec{\sigma}^{\mathcal{A}}\right) \| \tilde{\mathbf{P}}\left(-\vec{\sigma}^{\mathcal{A}}\right)\right) \geq  \ln\left(2\langle\boldsymbol{J}^{\mathcal{A}}\rangle^{T}{\Xi_{J^{\mathcal{A}}}^{-1}}\langle\boldsymbol{J}^{\mathcal{A}}\rangle+1\right)$ \, \cref{50}}}
\end{tabular}}}
\end{adjustbox}
\caption{The map between previously derived TURs and their MPP extensions.}
\label{tab:caption}
\end{table*}

\subsection{Extensions of previously derived TURs for scalar-valued currents to MPPs}
\label{sec:scalar.valued_currents_mpp}

In the following, we use the scalar-valued TURs summarized in \cref{sec:previous_TURs_review} to expand the local EPs in \cref{19}.

\subsubsection{Steady-state TUR}
\label{sec:ness_mpp}

Suppose that each unit $\omega$ is in a \textsc{NESS} as it evolves, though the individual steady states over each unit need not be identical. 
(Note that even though each unit is in an NESS, the joint system
need not be.) Then for each unit $\omega$,
\begin{align*}
\left<\sigma^{\omega}\right>\geq \frac{2\left<J^{\omega}\right>^{2}}{Var[J^{\omega}]} 
\tag{{\color{Black}{20}}} \label{20}
\end{align*}

We get, substituting \cref{20} into \cref{19},
\begin{align*}
\langle\sigma\rangle \geq \sum_{\omega \in \EuScript{N}_{o}^{*}} \frac{2\left\langle J^{\omega}\right\rangle^{2}}{\operatorname{Var}\left[J^{\omega}\right]}-\sum_{\omega \in \EuScript{N}_{e}^{*}}\left\langle\sigma^{\omega}\right\rangle
\tag{{\color{Black}{21}}} \label{21}
\end{align*}
that is applicable to both overlapping and non-overlapping units.

If none of the units overlap, then as there are no negative terms in the in-ex sum of local EPs, we can directly substitute \cref{20} into \cref{18}, and we get
\begin{align*}
\langle\sigma\rangle \geq \sum_{\omega \in \mathcal{N}^{*}} \frac{2\left\langle J^{\omega}\right\rangle^{2}}{V a r\left[J^{\omega}\right]}-\left\langle\Delta \mathcal{I}^{\mathcal{N}^{*}}\right\rangle
\tag{{\color{Black}{22}}} \label{22}
\end{align*}
\subsubsection{FTUR}
\label{sec:FTUR_mpp}

Next weaken the assumption that each unit is in an NESS,
to only only assume that the starting and ending distributions over each unit $\omega$ are identical, and that the driving protocols over each unit (which, once again, need not be identical) are time-symmetric. With these assumptions we can invoke \cref{3} to write
\begin{align*}
\left<\sigma^{\omega}\right>\geq \ln \left(\frac{2\left<J^{\omega}\right>^{2}}{Var[J^{\omega}]}+1\right) 
\tag{{\color{Black}{23}}} \label{23}
\end{align*}
for each unit $\omega$.
Plugging this expression in for the local EPs in \cref{19} gives
\begin{align*}
&\left<\sigma\right>\geq \sum_{\omega \in \EuScript{N}^{*}_{o}} \ln \left(\frac{2\left<J^{\omega}\right>^{2}}{Var[J^{\omega}]} + 1\right) -\sum_{\omega \in \EuScript{N}_{e}^{*}}\left\langle\sigma^{\omega}\right\rangle
\tag{{\color{Black}{24}}} \label{24}
\end{align*}
For non-overlapping units, alternatively, substitution of into \cref{18} would be used, giving us
\begin{align*}
\langle\sigma\rangle \geq \sum_{\omega \in \mathcal{N}^{*}} \ln \left(\frac{2\left<J^{\omega}\right>^{2}}{Var[J^{\omega}]} + 1\right) -\left\langle\Delta \mathcal{I}^{\mathcal{N}^{*}}\right\rangle
\tag{{\color{Black}{25}}} \label{25}
\end{align*}

\subsubsection{TUR involving instantaneous current}
\label{sec:instantaneous_mpp}

We noted in \cref{sec:intro} that there are also TURs 
that do not restrict either the initial or final distributions.
An example is given in 
\cite{6}. Define the instantaneous current over the joint system at time $t$ as $j_{t}(\boldsymbol{x})=\sum_{x \neq x^{\prime}} W_{x}^{x^{\prime}}(t) \mathbf{P}\left(\boldsymbol{x}_{t}^{\prime}\right) d_{x',x}$, and define $j^{\omega}_{t}(\textbf{x})$ to be any instantaneous current over $X_\omega$ at that time. Furthermore, introduce the time-integrated current over such unit as $J
^{\omega}(\textbf{x})=\int_{t_{\omega-1}}^{t_{\omega}} d t j^{\omega}_t(\textbf{x}).$ 
In \cite{6} it is shown that if the rate matrix for unit is time-homogeneous, then
even though
the overall system does not have to evolve time-homogeneously), it must be that:
\begin{align*}
\left<\sigma^{\omega}\right>\geq \frac{2\left<\tau^{\omega}
j^{\omega}_{t}\right>^{2}}{Var[J^{\omega}]}
\tag{{\color{Black}{26}}} \label{26}
\end{align*}

Substitution of \cref{26} to \cref{19} gives
\begin{align*}
\left<\sigma\right>\geq \sum_{\omega \in \EuScript{N}^{*}_{o}}\frac{2\left<\tau^{\omega} 
j^{\omega}_{t}\right>^{2}}{Var[J^{\omega}]} - \sum_{\omega \in \EuScript{N}_{e}^{*}}\left\langle\sigma^{\omega}\right\rangle
\tag{{\color{Black}{27}}} \label{27}
\end{align*}

In a parallel way, for non-overlapping units, substitution of \cref{26} to \cref{18} gives
\begin{align*}
\langle\sigma\rangle \geq \sum_{\omega \in \mathcal{N}^{*}} \frac{2\left<\tau^{\omega}
j^{\omega}_{t}\right>^{2}}{Var[J^{\omega}]}
-\left\langle\Delta \mathcal{I}^{\mathcal{N}^{*}}\right\rangle
\tag{{\color{Black}{28}}} \label{28}
\end{align*}

It is possible to repeat the procedure above to construct multipartite TURs on different levels --- n.b., this extension process would apply to any unit $\omega \in \mathcal{N}^{*}$ that would define its own unit structure under \textit{subsystem LDB}. (See Appendix D of \cite{22} for an inclusive discussion on SLDB.)
We could then consider any pair of units, $\omega$ and $\omega^{\prime}$, such that
\begin{align*}
\omega^{*}:=\left\{\omega^{\prime} \in \EuScript{N}^{*}: \omega^{\prime} \subseteq \omega\right\}
\tag{{\color{Black}{29}}} \label{29}
\end{align*}

Repeating the steps in \cref{sec:ness_mpp}, \cref{sec:FTUR_mpp}, \cref{sec:instantaneous_mpp}, we obtain \cref{30,31,32}. These expressions bound the local EPs of units in terms of the current precisions associated with their sub-units, with the corresponding change in the in-ex information over these sub-units. 
\begin{align*}
&\left<\sigma^{\omega}\right>\geq \sum_{\omega' \in {\omega^{*}_{o}}}  \frac{2\left<J^{\omega'}\right>^{2}}{Var[J^{\omega'}]}  - \sum_{\omega^{\prime}\in {\omega}_{e}^{*}}\left\langle\sigma^{\omega}\right\rangle
\tag{{\color{Black}{30}}} \label{30}
\end{align*}
\begin{align*}
&\left<\sigma^{\omega}\right>\geq \sum_{\omega' \in {\omega^{*}_{o}}} \ln \left(\frac{2\left<J^{\omega'}\right>^{2}}{Var[J^{\omega'}]} + 1\right) - \sum_{\omega^{\prime} \in {\omega}_{e}^{*}}\left\langle\sigma^{\omega}\right\rangle
\tag{{\color{Black}{31}}} \label{31}
\end{align*}
\begin{align*}
&\left<\sigma^{\omega}\right>\geq \sum_{\omega' \in {\omega^{*}_{o}}}\frac{2\left<\tau^{\omega'}
j^{\omega'}_{t}\right>^{2}}{Var[J^{\omega'}]} - \sum_{\omega^{\prime} \in {\omega}_{e}^{*}}\left\langle\sigma^{\omega}\right\rangle
\tag{{\color{Black}{32}}} \label{32}
\end{align*}

We remark that \cref{30} is a refinement to \cref{21}, where \cref{31} is to \cref{24}, and \cref{32} is to \cref{27}.

\subsubsection{Mixed multipartite TURs} 

The multipartite structure allows that each current precision term, computed for distinct units that are driven with differing physical procedures would still contribute to the lower bound on the average EP of the global system that these units compose, even if the global system doesn’t obey \textit{any} TUR. Consider for instance a unit structure that is composed of three distinct units $A$, $B$, $C$ that evolve under differing drifing protocols, where A's assumed to be in a NESS (i.e., A's driving protocol is strictly time-independent), B evolves under a time-symmetric protocol, and C evolves so that its distribution at $t = 0$ and $t=t_{f}$ are identical. For such a setting, we would express the ``mixed TUR'' that includes the local current precisions and the global EP as:
\begin{align*}
&\left<\sigma\right>\geq \frac{2\left\langle J^{A}\right\rangle^{2}}{\operatorname{Var}\left[J^{A}\right]} + \ln \left(\frac{2\left<J^{B}\right>^{2}}{Var[J^{B}]} + 1\right) + \frac{2\left<\tau^{C}
j^{C}_{t}\right>^{2}}{Var[J^{C}]} 
\tag{{\color{Black}{33}}} \label{33}
\end{align*}
\label{sec:instantaneous_mpp}
\subsection{Conditional TURs based on DFTs for multipartite processes}
\label{sec:conditional.TUR_mpp}

In \cite{22} a conditional DFT is derived, applicable for any unit $\omega$,
and any associated EP value $\sigma^{\omega}$ that has non-zero probability:
\begin{align*}
\left\langle\sigma \mid \sigma^{\omega}\right\rangle \geq \sigma^{\omega}
\tag{{\color{Black}{34}}} \label{34}
\end{align*}
\cref{34} means that the expected global EP conditioned on the joint EP of $\mathcal{A}$ and  $\mathcal{B}$ cannot be smaller than the joint EP of $\mathcal{A}$ and $\mathcal{B}$. It also means that the expected joint EP of $\mathcal{A}$ and $\mathcal{B}$ conditioned on the EP of $\mathcal{B}$ cannot be smaller than the EP of $\mathcal{B}$. 

Note that the bound in \cref{34} is saturated
if the value $\sigma^{\omega}$ happens to equal the expected EP of unit $\omega$,
$\langle \sigma^{\omega} \rangle$ (assuming that can occur with non-zero probability). For this particular case where the observed EP equals $\langle \sigma^{\omega} \rangle$,
we can plug in the results of the previous sections into the RHS of \cref{34}, to
derive the following conditional TURs for MPPs:
\begin{align*}
&\left\langle\sigma \mid \sigma^{\omega}\right\rangle \geq \sum_{\omega' \in \omega^{*}_{o}} \frac{2\left<J^{\omega'}\right>^{2}}{Var[J^{\omega'}]} - \sum_{\omega^{\prime}\in {\omega}_{e}^{*}}\left\langle\sigma^{\omega}\right\rangle \tag{{\color{Black}{35}}} \label{35} \\
&\left\langle\sigma \mid \sigma^{\omega}\right\rangle\geq \sum_{\omega' \in \omega^{*}_{o}} \ln \left(\frac{2\left<J^{\omega'}\right>^{2}}{Var[J^{\omega'}]} + 1\right) - \sum_{\omega^{\prime}\in {\omega}_{e}^{*}}\left\langle\sigma^{\omega}\right\rangle \tag{{\color{Black}{36}}} \label{36} \\
&\left\langle\sigma \mid \sigma^{\omega}\right\rangle \geq \sum_{\omega' \in \omega^{*}_{o}}\frac{2\left<\tau^{\omega'}
j^{\omega'}_{t}\right>^{2}}{Var[J^{\omega'}]} - \sum_{\omega^{\prime}\in {\omega}_{e}^{*}}\left\langle\sigma^{\omega}\right\rangle \tag{{\color{Black}{37}}} \label{37}
\end{align*}

\subsection{Extensions of previously derived TURs for vector-valued currents to MPPs}
\label{sec:vector.valued_currents_MPP}

In the analysis of the preceding sections, there was only one current specified for each
unit, i.e., each unit $\omega$ had a single associated increment function $d_\omega(., .)$. In
general though, any subsystem will have many possible associated increment functions, and the associated currents can be statistically coupled. This
results in TURs for ``vector-valued" current. In this section we present several
results related to such current.

To begin, we  review
the vector-valued fluctuation theorem for MPPs derived
in \cite{22}, and a joint fluctuation theorem of EP and current over a single system derived in \cite{5}. We then combine these results to derive FTURs 
for vector-valued currents that apply to MPPs. 

Our first results apply for any pairs of \textit{sets} of units, $\mathcal{A}$, $\mathcal{B}$,
which in general may or may not have subsystems in common. Each of those sets has an
associated vector of currents, indexed by the units in that set, which we write as
$\vec{J}^{\mathcal{A}}$ and $\vec{J}^{\mathcal{B}}$, respectively. Note that when the sets do
indeed overlap, there is statistical coupling between $\vec{J}^{\mathcal{A}}$ and $\vec{J}^{\mathcal{B}}$.

We derive a bound on the precisions of both $\vec{J}^{\mathcal{A}}$ and $\vec{J}^{\mathcal{B}}$. This bound depends on the covariance of those current vectors, and on the EPs
within $\mathcal{A}$ and $\mathcal{B}$ --- but does not explicitly involve the global EP. As a special case, we present a bound on
the local precision of a vector-valued current over the subsystems within a single unit (i.e., 
a bound on the \textit{overall} EP generated in that unit).

Vector-valued currents have already been analyzed previously for the special
case of a single system in \cite{8}. We next show how to extend
these results to MPPs,
concentrating on a few specific scenarios.

Special cases of some of our findings in this section concern scenarios that were already addressed by previous
results in the literature. Our findings do not necessarily tighten such earlier results. However, 
our results extend beyond those special cases, to tease apart some of the different 
bounds constraining how the thermodynamic properties of different subsystems 
can interact with one another. These bounds provide us several trade-off relations concerning the thermodynamics of MPPs.

\subsubsection{The multi-variable FTUR for multipartite processes} 
\label{sec:vector.FTUR_mpp}

Let $\mathcal{A} = \{\alpha\}$ be any set of units. We write
$\vec{\sigma}^{A}$ for the associated vector whose components are the local EP values $\sigma^{\alpha}$, and $\vec{J}^{\mathcal{A}}$ for the 
associated vector whose components are the individual currents ${J}^{\alpha}$. We also
write $\sigma^{\mathcal{A}}(\boldsymbol{x})$ for the joint EP generated by all the subsystems in $\mathcal{A}$, while $J^{\mathcal{A}}(\boldsymbol{x})$ is the total current generated by the subsystems in $\mathcal{A}$. 

It was shown in \cite{22} that the following vector-valued fluctuation theorem holds:
\begin{align*}\ln \left[\frac{\mathbf{P}\left(\vec{\sigma}^{ \mathcal{A}}\right)}{\tilde{\mathbf{P}}\left(-\vec{\sigma}^{\mathcal{A}}\right)}\right]=\sigma^{\cup \mathcal{A}}
\tag{{\color{Black}{38}}} \label{38}
\end{align*}
In Appendix B we show that in fact a vector-valued joint fluctuation theorem over both
local EPs and local currents holds:
\begin{align*}
\ln \left[\frac{\mathbf{P}\left(\vec{\sigma}^{\mathcal{A}},\vec{J}^{\mathcal{A}}\right)}{\tilde{\mathbf{P}}\left(-\vec{\sigma}^{\mathcal{A}},-\vec{J}^{ \mathcal{A}}\right)}\right]=\sigma^{\cup \mathcal{A}}
\tag{{\color{Black}{39}}} \label{39}
\end{align*}
We can use \cref{39} to investigate the statistical dependencies between the local currents generated by different units. Recall that for any set of subsystems $\mathcal{A}$,
$\vec{J}^{\mathcal{A}}$ indicates the vector whose components are the individual currents ${J}^{\alpha}$ for units $\alpha \in \mathcal{A}$.
Similarly, for any set of subsystems $\mathcal{B}$,
$\vec{J}^{\mathcal{B}}$ indicates the vector whose components are the individual currents ${J}^{\beta}$ for units $\beta \in \mathcal{B}$.

If $\mathcal{A}$ and $\mathcal{B}$ overlap, then the vector-valued currents 
$\vec{J}^{\mathcal{A}}$ and $\vec{J}^{\mathcal{B}}$ will be statistically
coupled, in general.
In Appendix D we derive the multi-variable FTUR for MPPs for any pairs of units $\alpha_{i}$, $\beta_{i}$,
\begin{align*}
\frac{\operatorname{Covar}\left[J^{\alpha_{i}}, J^{\beta_{j}}\right]}{\left\langle J^{\alpha_{i}}\right\rangle\left\langle J^{\beta_{j}}\right\rangle}\geq \frac{2 \cdot \left(1+e^{\langle\sigma^{\cup \mathcal{A}}-\sigma^{\cup \mathcal{\mathcal{B}}}\rangle}\right)}{e^{\langle\sigma^{\cup \mathcal{\mathcal{A}}}\rangle}+e^{\langle\sigma^{\cup \mathcal{\mathcal{B}}}\rangle}-\left(1+e^{\langle\sigma^{\cup \mathcal{\mathcal{A}}}-\sigma^{\cup \mathcal{\mathcal{B}}}\rangle}\right)}
\tag{{\color{Black}{41}}} \label{41}
\end{align*}
where $\sigma^{\cup \mathcal{A}}$ is the total EP generated by the subsystems in $\mathcal{A}$,
and $\sigma^{\cup \mathcal{B}}$ is defined similarly.
\cref{41} holds in any setting where the driving protocol is time-symmetric. Note that when $\mathcal{A} = \mathcal{B}$, \cref{41} reduces to
\begin{align*}
\frac{\operatorname{Var} J^{\alpha_{i}}}{\left\langle J^{\alpha_{i}}\right\rangle^{2}} \geq \frac{2}{e^{\langle\sigma^{\cup \mathcal{A}}\rangle}-1}
\tag{{\color{Black}{42}}} \label{42}
\end{align*}
(In Appendix C we derive \cref{42} in a more straight-forward manner,
by invoking \cref{39} explicitly.) \cref{42} shows how $\sigma^{\cup \mathcal{A}}$ bounds the precisions of 
the local currents.

\subsubsection{MTUR for multipartite processes}
\label{sec:dechant_mpp}

In \cref{sec:vector.FTUR_mpp}, we used FTs for MPPs to obtain multi-variable TURs for MPPs.
In this subsection we instead derive multi-variable TURs for MPPs by building on
the analysis in \cite{8}.

We start with a bound
derived in \cite{8}:
\begin{align*}
\langle\boldsymbol{J}\rangle^{T} \Xi_{J}^{-1}\langle\boldsymbol{J}\rangle \leq \frac{1}{2} \Delta S
\tag{{\color{Black}{43}}} \label{43}
\end{align*}

Consider each component of $\boldsymbol{J}$ in \cref{43} as a local current originated in a single unit. Suppose also that the inverse covariance matrix is diagonal, implying that the units are independent. Given these conditions, we note that LHS equals the sum of the precisions of the currents within individual units. Substituting \cref{20} to the RHS of \cref{43} gives
\begin{align*}
\langle\boldsymbol{J}\rangle^{T} \Xi_{J}^{-1}\langle\boldsymbol{J}\rangle \leq \frac{\sum_{w} \sigma^{\omega}(\textit{\textbf{x}}_\omega)-\Delta \mathcal{I}^{{\mathcal{N}^*}}(\textit{\textbf{x}})}{2}
\tag{{\color{Black}{44}}} \label{44}
\end{align*}

If the units overlap, 
but the change in in-ex information
is positive, then the LHS of \cref{44} would be upper-bounded by the in-ex sum of local EPs that are generated in distinct units. Alternatively, if the in-ex sum of local EPs is non-negative, then the LHS of \cref{44} would be upper-bounded by the drop in the in-ex information.

Following \cref{44}, we also present \cref{45} in a parallel form to \cref{3}:
\begin{align*}
\langle\boldsymbol{J}^{\mathcal{A}}\rangle^{T}{\Xi_{J^{\mathcal{A}}}^{-1}}\langle\boldsymbol{J}^{\mathcal{A}}\rangle \leq \frac{e^{\langle\sigma^{\cup \mathcal{A}}\rangle}-1}{2}
\tag{{\color{Black}{45}}} \label{45}
\end{align*}
Equivalently,
\begin{align*}
\langle\sigma^{\cup \mathcal{A}}\rangle \geq  \ln\left(2\langle\boldsymbol{J}^{\mathcal{A}}\rangle^{T}{\Xi_{J^{\mathcal{A}}}^{-1}}\langle\boldsymbol{J}^{\mathcal{A}}\rangle+1\right) 
\tag{{\color{Black}{46}}} \label{46}
\end{align*}
In addition, for the global system, we write
\begin{align*}
\langle\sigma\rangle \geq  \ln\left(2\langle\boldsymbol{J}\rangle^{T}{\Xi_{J}^{-1}}\langle\boldsymbol{J}\rangle+1\right)
\tag{{\color{Black}{47}}} \label{47} 
\end{align*}
Note that if we consider a scalar current only, then this expression boils down to:
\begin{align*}
\langle\sigma\rangle \geq  \ln\left(\sum_{\omega \in \EuScript{N^*}} {{\frac{2\left\langle J^{\omega} \right\rangle^{2}}{Var[J^{\omega}]}}}+1\right) 
\tag{{\color{Black}{48}}} \label{48}
\end{align*}

Also, note that we can express \cref{43} and \cref{44} in a different form, by using a finding from \cite{22} given in \cref{49}, which states that the expected EP over a set of subsystems is given by the KL-divergence between the (forward and backward) probability distributions of the EP vector over such subsystems, where each component of the EP vector is mapped to a scalar local EP value over one unit that is a part of the overall unit structure.
\begin{align*}
\left\langle\sigma^{\cup \mathcal{A}}\right\rangle=D\left(\mathbf{P}\left(\vec{\sigma}^{\mathcal{A}}\right) \| \tilde{\mathbf{P}}\left(-\vec{\sigma}^{\mathcal{A}}\right)\right)
\tag{{\color{Black}{49}}} \label{49}
\end{align*}
Substituting \cref{49} to \cref{44}, and to \cref{45} would then give, respectively, \cref{50} and \cref{51}:
\begin{align*}
&D\left(\mathbf{P}\left(\vec{\sigma}^{\mathcal{A}}\right) \| \tilde{\mathbf{P}}\left(-\vec{\sigma}^{\mathcal{A}}\right)\right) \geq  2\langle\boldsymbol{J}^{\mathcal{A}}\rangle^{T}{\Xi_{J^{\mathcal{A}}}^{-1}}\langle\boldsymbol{J}^{\mathcal{A}}\rangle
\tag{{\color{Black}{50}}} \label{50} \\
&D\left(\mathbf{P}\left(\vec{\sigma}^{\mathcal{A}}\right) \| \tilde{\mathbf{P}}\left(-\vec{\sigma}^{\mathcal{A}}\right)\right) \geq  \ln\left(2\langle\boldsymbol{J}^{\mathcal{A}}\rangle^{T}{\Xi_{J^{\mathcal{A}}}^{-1}}\langle\boldsymbol{J}^{\mathcal{A}}\rangle+1\right) 
\tag{{\color{Black}{51}}} \label{51}
\end{align*}

\section{EXAMPLES}
\label{sec:examples}

Here we illustrate our findings in a setting of multiple single-level quantum dots coupled to multiple reservoirs. The particular physical scenarios we consider are depicted in Fig.\,4.

\begin{figure}[h!]
\centering
     \begin{subfigure}{0.44\textwidth}
       \centering
       \includegraphics[scale=0.4]{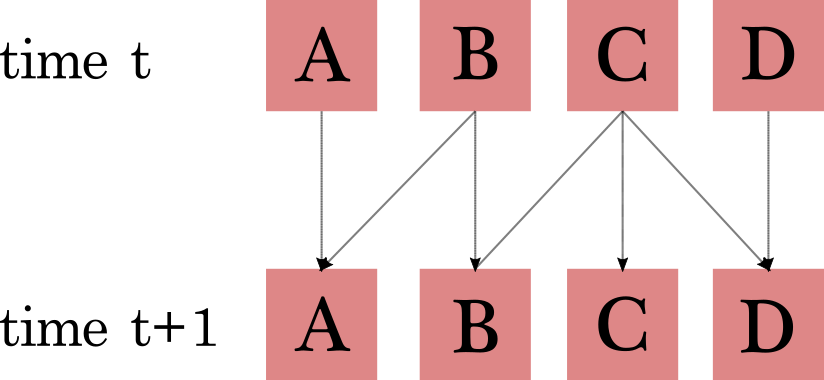}
       \caption{Four single-level quantum dots that comprise three overlapping units.}\label{fig:a}
    \end{subfigure}          
    \qquad\qquad % spacing between the subfigures
    \begin{subfigure}{0.44\textwidth}  
       \centering
       \includegraphics[scale=0.4]{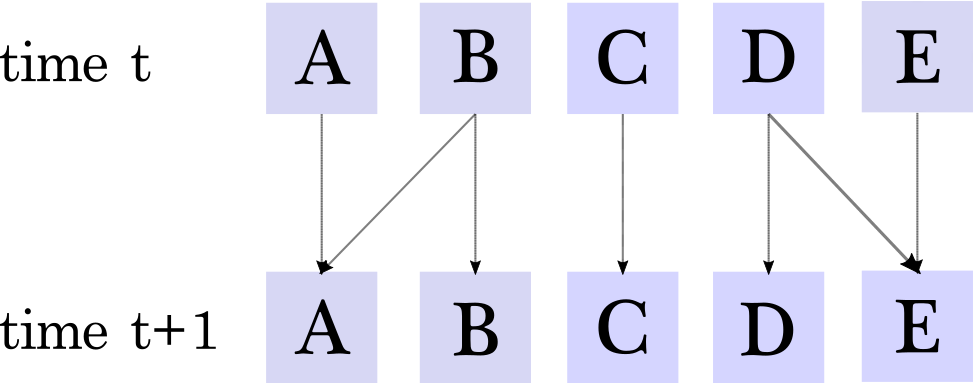}
       \caption{Five single-level quantum dots that comprise three non-overlapping units.}\label{fig:aa}
    \end{subfigure}    
    \caption{Settings in which the multipartite TURs are numerically investigated.}
\end{figure}

We follow the analysis in \cite{26}, and obtain the rate equations for multiple quantum dots through an extension of Eq. (97) in \cite{26}, which presents the explicit full dynamics of double quantum dots. The quantum dots are identified by states up $u$ and down $d$ with fixed energies $\varepsilon_{u}$ and $\varepsilon_{d}$, and 
each dot $n$ is coupled with a reservoir of temperature $T_n$ and chemical potential $\mu_n$. For the NESS, we assume for both of the settings depicted in Fig.\,4, all the dots (i.e., all subsystems $A-D$ and $A-E$) are coupled to distinct multiple reservoirs. These reservoirs are assumed to have the same temperature and different chemical potentials. 
In the most general form, the rate matrix $W_{x}^{x^{\prime}}$ for the scenario depicted in Fig.\,4(a) is
\begin{align*}
&W_{x}^{x^{\prime}}(\{A,B,C\} ; t)= W_{x_{A}, x_{B}, x_{C}}^{x_{A}^{\prime}, x_{B}^{\prime}, x_{C}^{\prime}}(A, t)\\
& \hspace{2.72cm}+ W_{x_{A}, x_{B}, x_{C}}^{x_{A}^{\prime}, x_{B}^{\prime}, x_{C}^{\prime}}(B, t) \\
&  \hspace{2.72cm}+ W_{x_{A}, x_{B}, x_{C}}^{x_{A}^{\prime},  x_{B}^{\prime}, x_{C}^{\prime}}(C, t) \\
&W_{x}^{x^{\prime}}(\{C,D\} ; t)= W_{x_{C}, x_{D}}^{x_{C}^{\prime}, x_{D}^{\prime}}(C, t)+W_{x_{C}, x_{D}}^{x_{C}^{\prime}, x_{D}^{\prime}}(D, t)
\\
&W_{x}^{x^{\prime}}(\{C\} ; t)=W_{x_{C}}^{x_{C}^{\prime}}(C, t)
\tag{{\color{Black}{52}}} \label{52}
\end{align*}
and the rate matrix $\hat{W}_{x}^{x^{\prime}}$ for the scenario depicted in Fig.\,4(b) is
\begin{align*}
&\hat{W}_{x}^{x^{\prime}}(\{A,B\} ; t)=\hat{W}_{x_{A}, x_{B}}^{x_{A}^{\prime}, x_{B}^{\prime}}(A, t)+\hat{W}_{x_{A}, x_{B}}^{x_{A}^{\prime}, x_{B}^{\prime}}(B, t)\\
&\hat{W}_{x}^{x^{\prime}}(\{D,E\} ; t)=\hat{W}_{x_{D}, x_{E}}^{x_{D}^{\prime}, x_{E}^{\prime}}(D, t)+\hat{W}_{x_{D}, x_{E}}^{x_{D}^{\prime}, x_{E}^{\prime}}(E, t)
\\
&\hat{W}_{x}^{x^{\prime}}(\{C\} ; t)=\hat{W}_{x_{C}}^{x_{C}^{\prime}}(C, t)
\tag{{\color{Black}{53}}} \label{53}
\end{align*}

We introduce the driving term in the rate equations by parametrizing a time-dependence of the energy of a single dot $\epsilon$, for each subsystem through an external driving term $\lambda$. 
(The rates that characterize each reservoir-dot coupling are now written in terms of an effective potential, $\hat{\epsilon}(t)=\epsilon(t)-\mu$, i.e., $W_{ij}(t)=1/e^{\beta \hat{\epsilon}(t)}+1$.) \cref{52} and \cref{53} satisfy the GTUR when the initial and final states are in the steady state and $\lambda(t)=$ $\lambda(T-t)$, i.e., the system is driven by a time-symmetric protocol. To illustrate this set of scenarios, we choose the energy gap between $u$ and $d$ as  $\Delta = 1 - \alpha$ where $\alpha$ is a control parameter. In the experiments
illustrated in Fig.\,4(a), all reservoirs have $\beta = 1,$ $\alpha = 0.4$, and to make Fig.\,4(b) $\beta = 2,$ $\alpha = 0.3$. The steady state for both settings is given by $ \dot{p}_{x}=\sum_{x^{\prime}} W_{x x^{\prime}} p_{x^\prime} = 0.$ 

In the following, we define $X^{-1} := \frac{\left<J^{\omega}\right>^{2}}{Var[J^{\omega}]}$, 
which we extend to write $X^{-1}_{\omega}$ to indicate $X^{-1}$ computed for the specific unit $\omega$. Similarly, we define $Y^{-1} = \ln \left(\frac{2\left<J^{\omega}\right>^{2}}{Var[J^{\omega}]} + 1\right)$, and $Z^{-1} =\frac{2\left<\tau^{\omega}
j^{\omega}_{t}\right>^{2}}{Var[J^{\omega}]}$. The sums in the bounds given by \cref{22}, \cref{25} and \cref{28} is written as \textsf{$\sum$} in the legends. 

Fig.\,5(a) illustrates \cref{22} for an NESS, for the scenario of three non-overlapping units, denoted by $\omega$, $\omega^{\prime}$, and $\omega^{\prime \prime}$. Next, Fig.\,5(b) shows how the reciprocal form of \cref{22} can be used to bound the global EP. Fig.\,6(a) and Fig.\,6(b) are analogous to Fig.\,5(a) and Fig.\,6(b), respectively, but for the overlapping units scenario illustrated in Fig.\,4(a). Fig.\,7 addresses a multipartite TUR, governed by \cref{33}, where three non-overlapping units evolve according to three different driving protocols that allow each local current precision to be bounded by $X$, $Y$, and $Z$, respectively. Fig.\,8 compares the tightness of the multipartite NESS TUR given by \cref{22}, the multipartite FTUR given by \cref{25}, and the multipartite TUR for instantaneous currents given by \cref{28}. The curves in Fig.\,8 imply that in an NESS generated by three non-overlapping units, the bound given by the multipartite TUR for instantaneous currents is closest to global mean dissipation.

Our findings also reflect the trade-offs between various units in bounding the global EP, which is captured by the change in the in-ex information. Fig.\,9 demonstrates this through a scenario where the global current over a joint system is fixed.
\begin{figure}[h]
    \centering
     \begin{subfigure}{0.48\textwidth}
       \centering
       \includegraphics[scale=0.128]{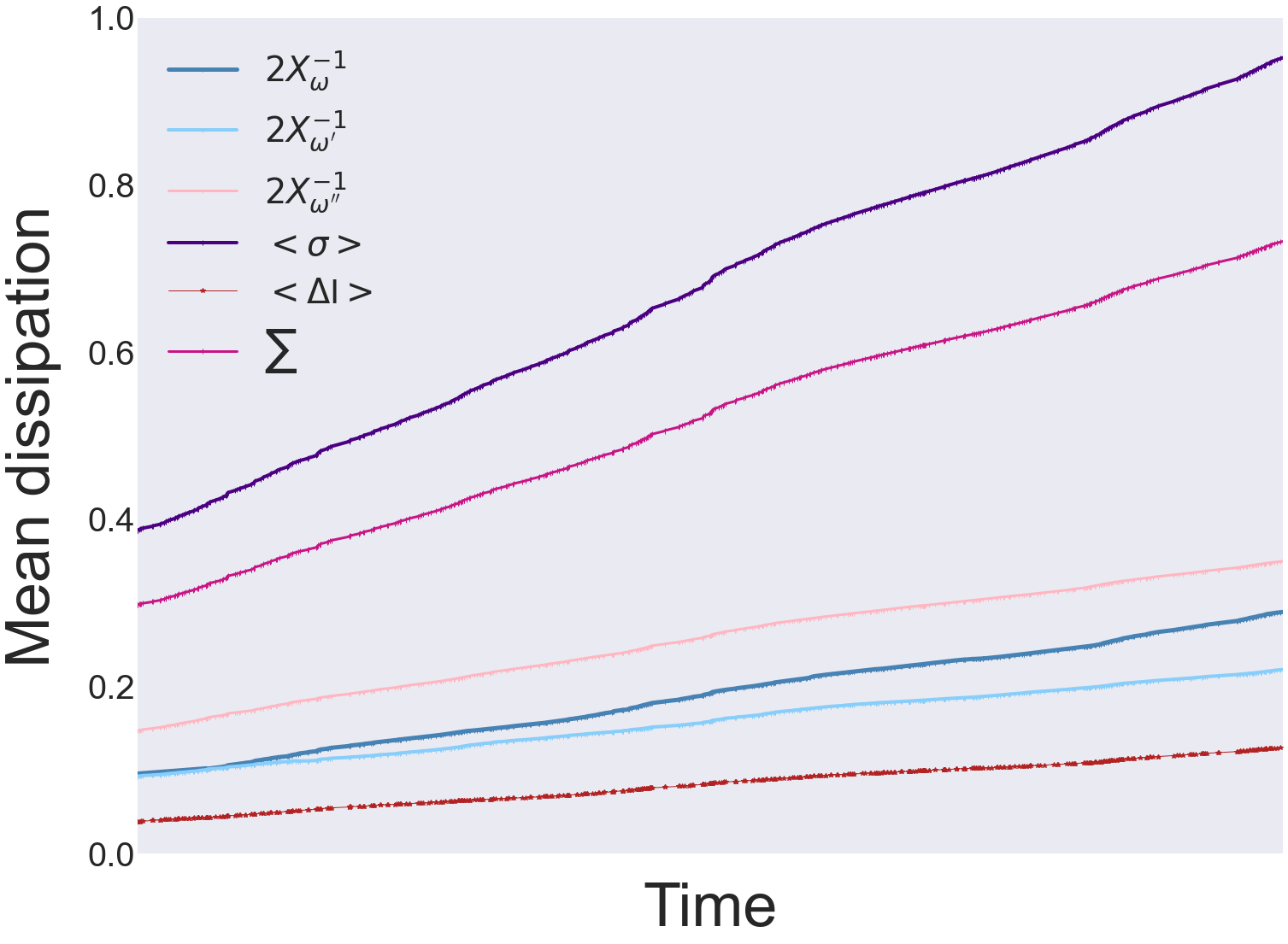}
       \caption{Mean dissipation versus time graph for the setting captured by Fig.\,4(b).}\label{fig:a}
    \end{subfigure}          
    \,
    \begin{subfigure}{0.48\textwidth}  
       \centering
       \includegraphics[scale=0.128]{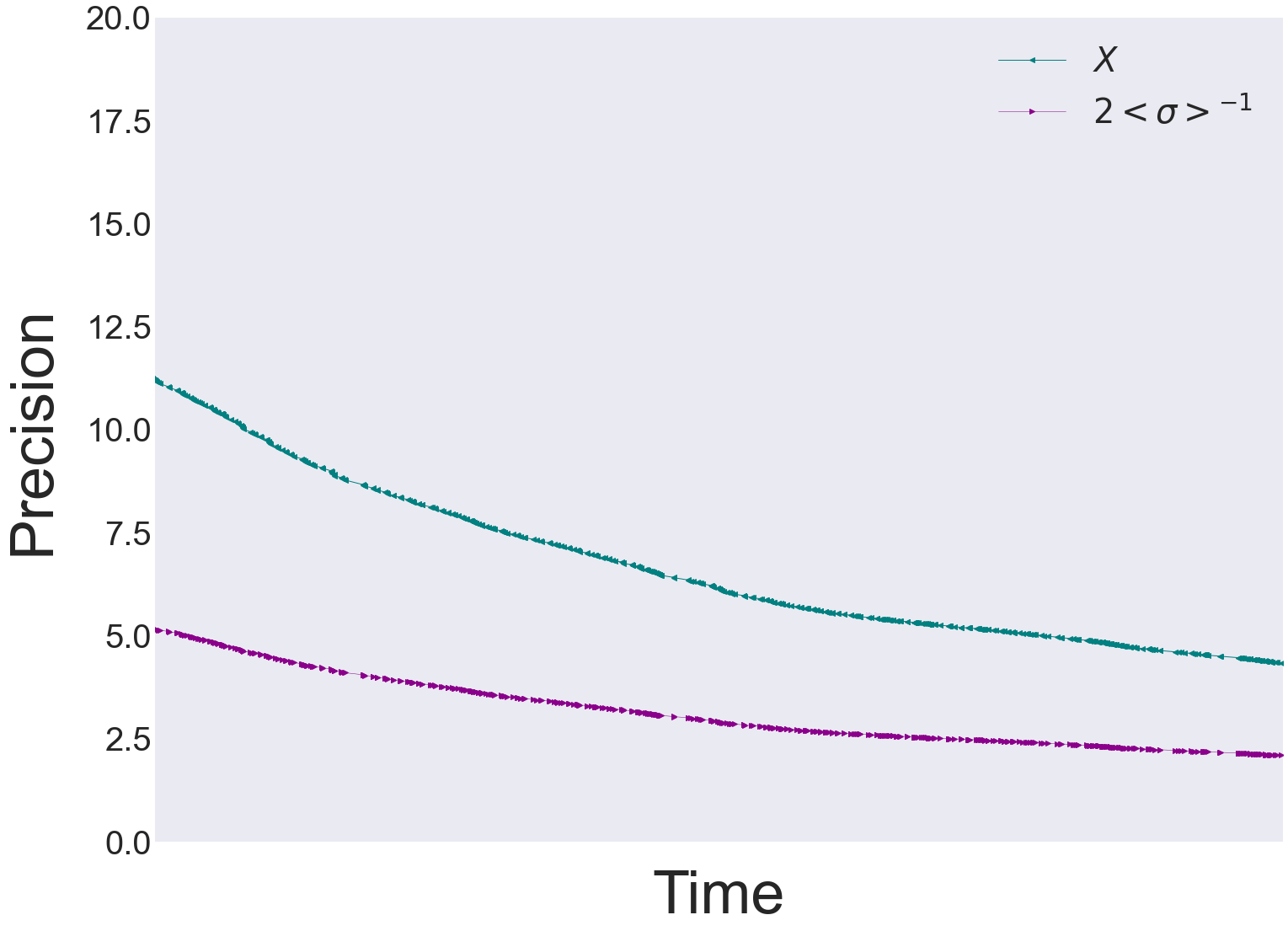}
       \caption{Precision and inverse-mean dissipation for the same setting as Fig.\,5(a).}\label{fig:aa}
    \end{subfigure}    
    \caption{For (a), the components that compose \cref{22} are given on the \textit{y}-axis. $2X^{-1}_{\omega}$, $2X^{-1}_ {\omega^{\prime}},2X^{-1}_ {\omega^{\prime\prime}} $ denote the contributions from all three units to the in-ex sum on the RHS of \cref{22}.}
\end{figure}

\begin{figure}[h!]
    \centering
     \begin{subfigure}{0.48\textwidth}
       \centering
       \includegraphics[scale=0.128]{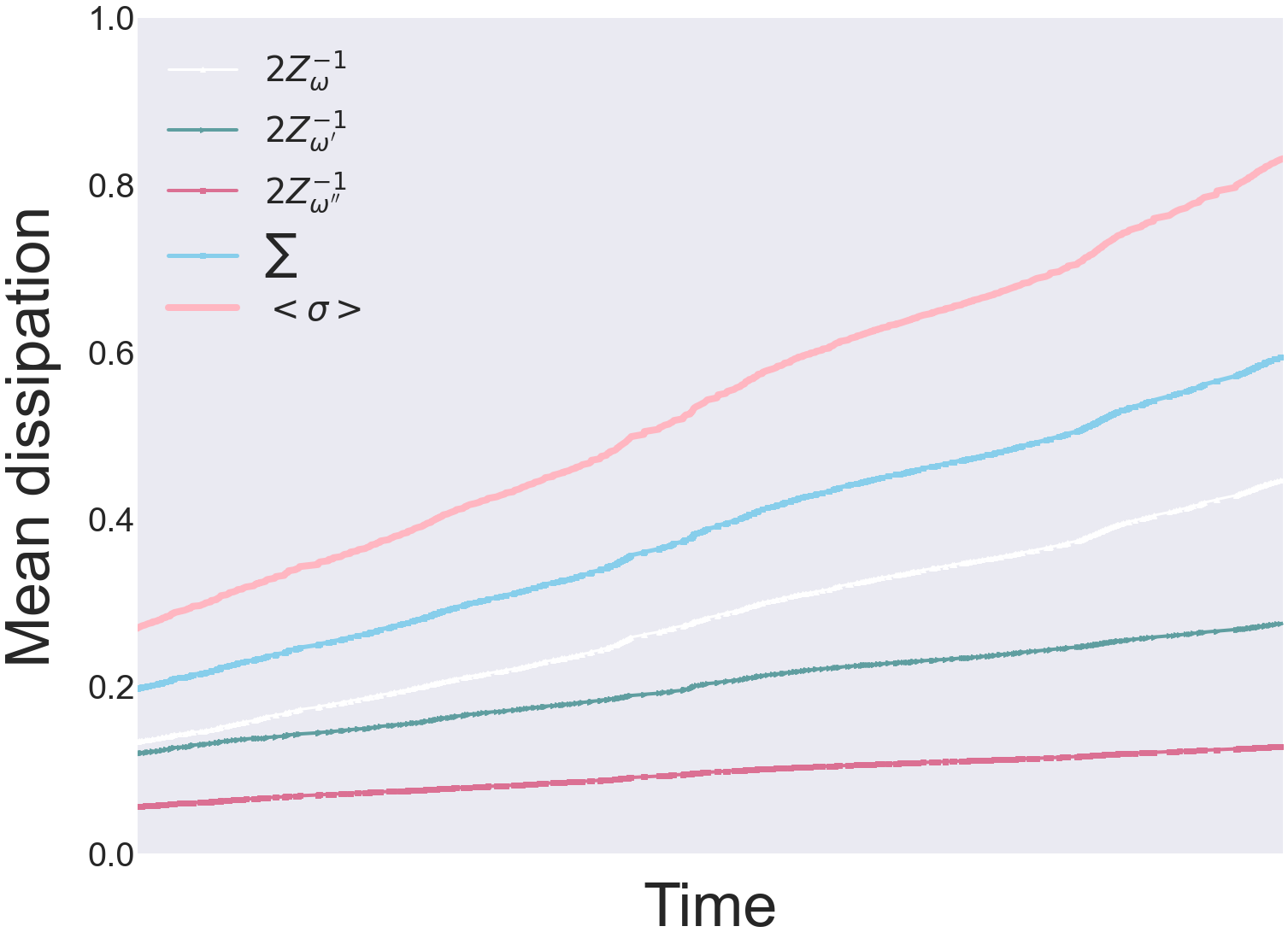}
       \caption{Mean dissipation versus time graph for the setting captured by Fig.\,4(a).}\label{fig:a}
    \end{subfigure}          
    \,
    \begin{subfigure}{0.48\textwidth}  
       \centering
       \includegraphics[scale=0.128]{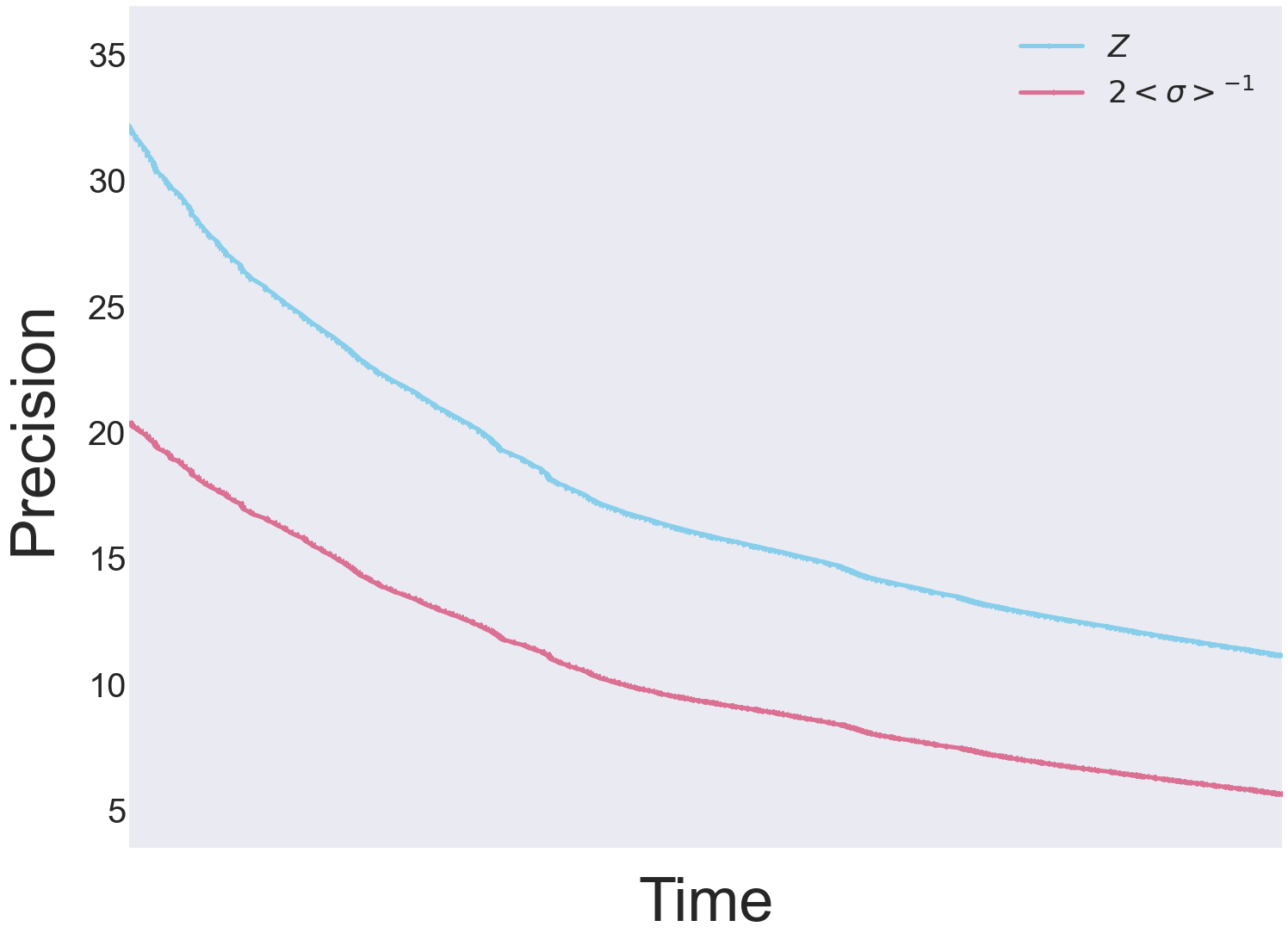}
       \caption{Precision and inverse-mean dissipation graph for the same setting as Fig.\,6(a).}\label{fig:aa}
    \end{subfigure}    
    \caption{For (a), the components that compose \cref{27} are given on the \textit{y}-axis. $2X^{-1}_{\omega}$, $2X^{-1}_ {\omega^{\prime}},2X^{-1}_ {\omega^{\prime\prime}} $ denote the contributions from all three units to the in-ex sum on the RHS of \cref{27}.}
\end{figure}
\begin{figure}[h!]
    \centering
     \begin{subfigure}{0.44\textwidth}
       \centering
       \includegraphics[scale=0.128]{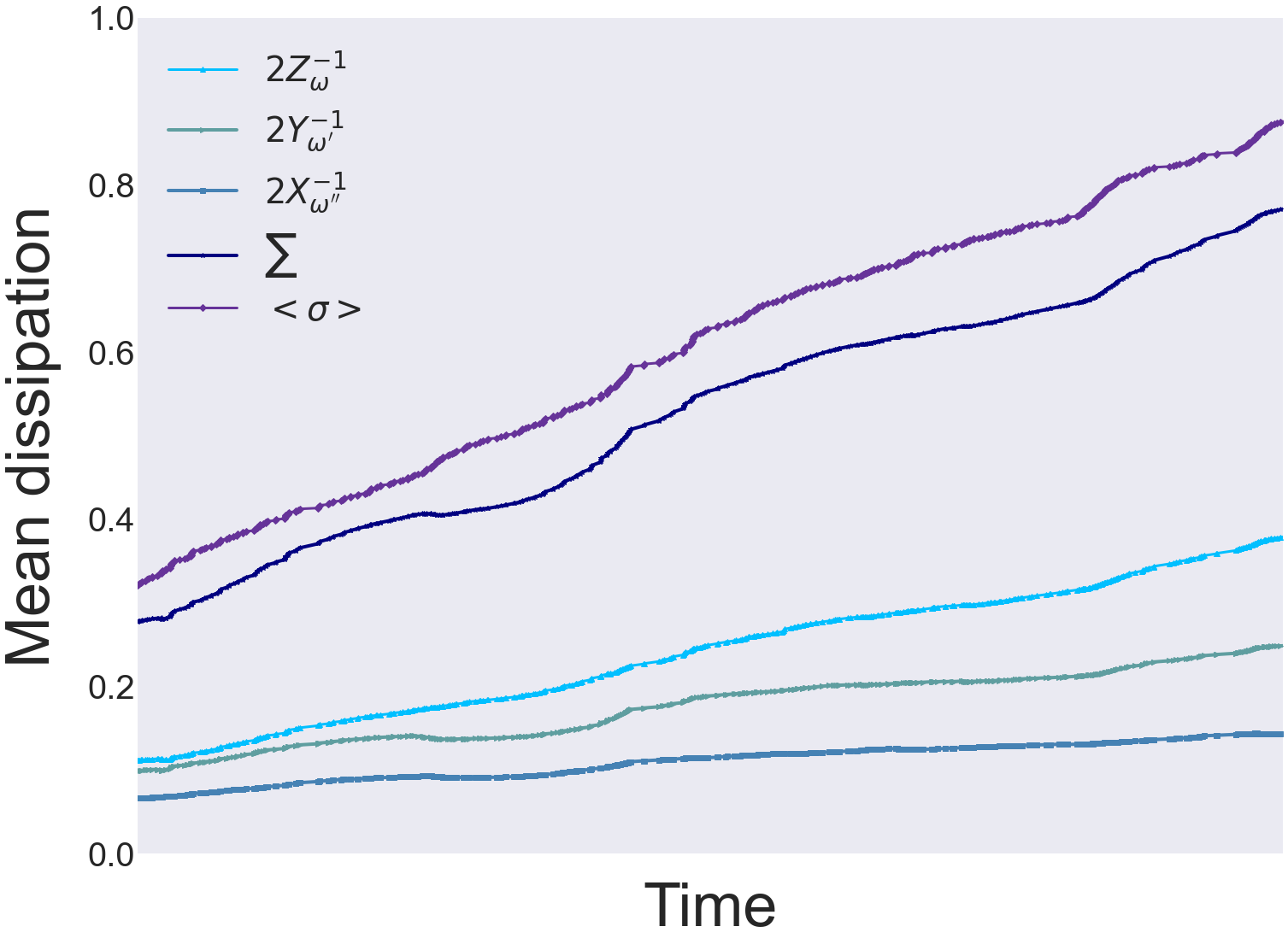}
       \caption{Mean dissipation versus time graph for the setting captured by Fig.\,4(b).}\label{fig:a}
    \end{subfigure}          
    \quad
    \begin{subfigure}{0.44\textwidth}  
       \centering
       \includegraphics[scale=0.128]{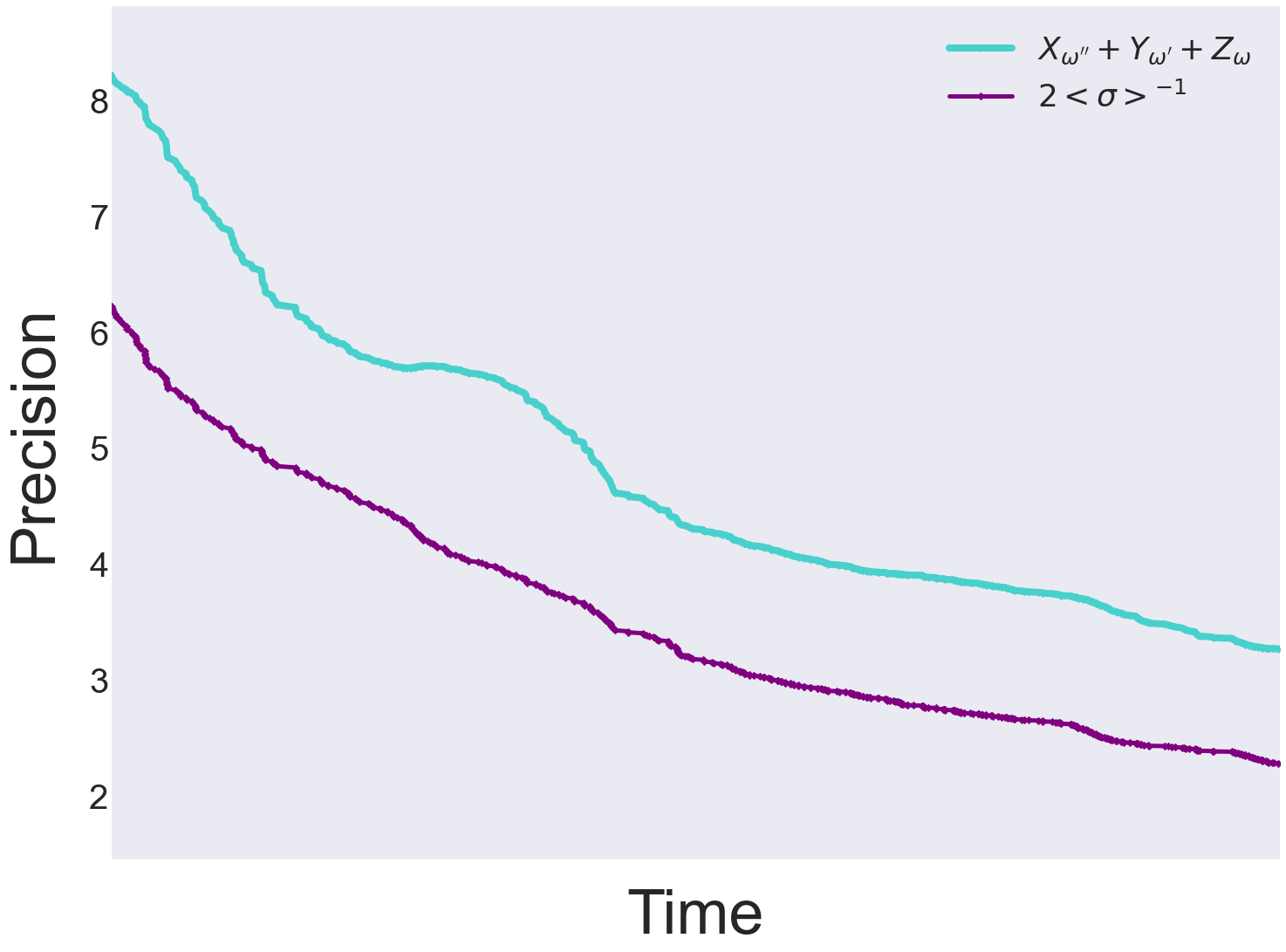}
       \caption{Precision and inverse-mean dissipation graph for the same setting as Fig.\,7 (a).}\label{fig:aa}
    \end{subfigure}    
    \caption{For (a), the separate terms in \cref{33} are given on the \textit{y}-axis. $2Z^{-1}_{\omega}$, $2Y^{-1}_ {\omega^{\prime}},2X^{-1}_ {\omega^{\prime\prime}} $ are the contributions from all three units to the in-ex sum on the RHS of \cref{33}.}
\end{figure}
\begin{figure}[h]
\centering
\begin{minipage}{.44\textwidth}
  \centering
  \includegraphics[width=1.00\linewidth]{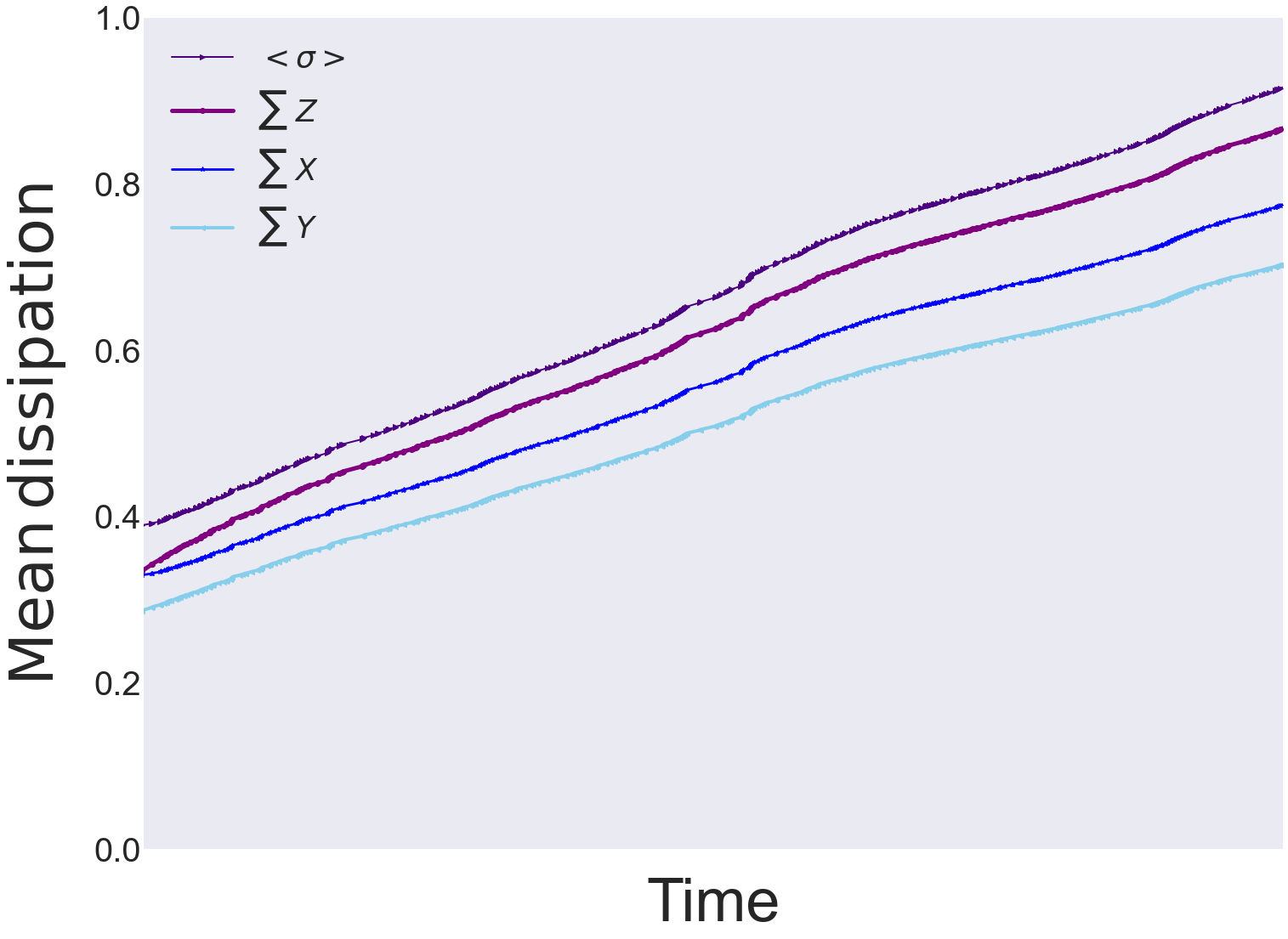}
  \captionof{figure}{Mean dissipation and relevant in-ex sums, computed for Eq. (22), Eq. (25), and Eq. (28), labeled by $X$, $Y$, and $Z$, respectively.}
  \label{fig:test1}
\end{minipage} \quad
\begin{minipage}{.48\textwidth}
  \centering
  \includegraphics[width=0.66\linewidth]{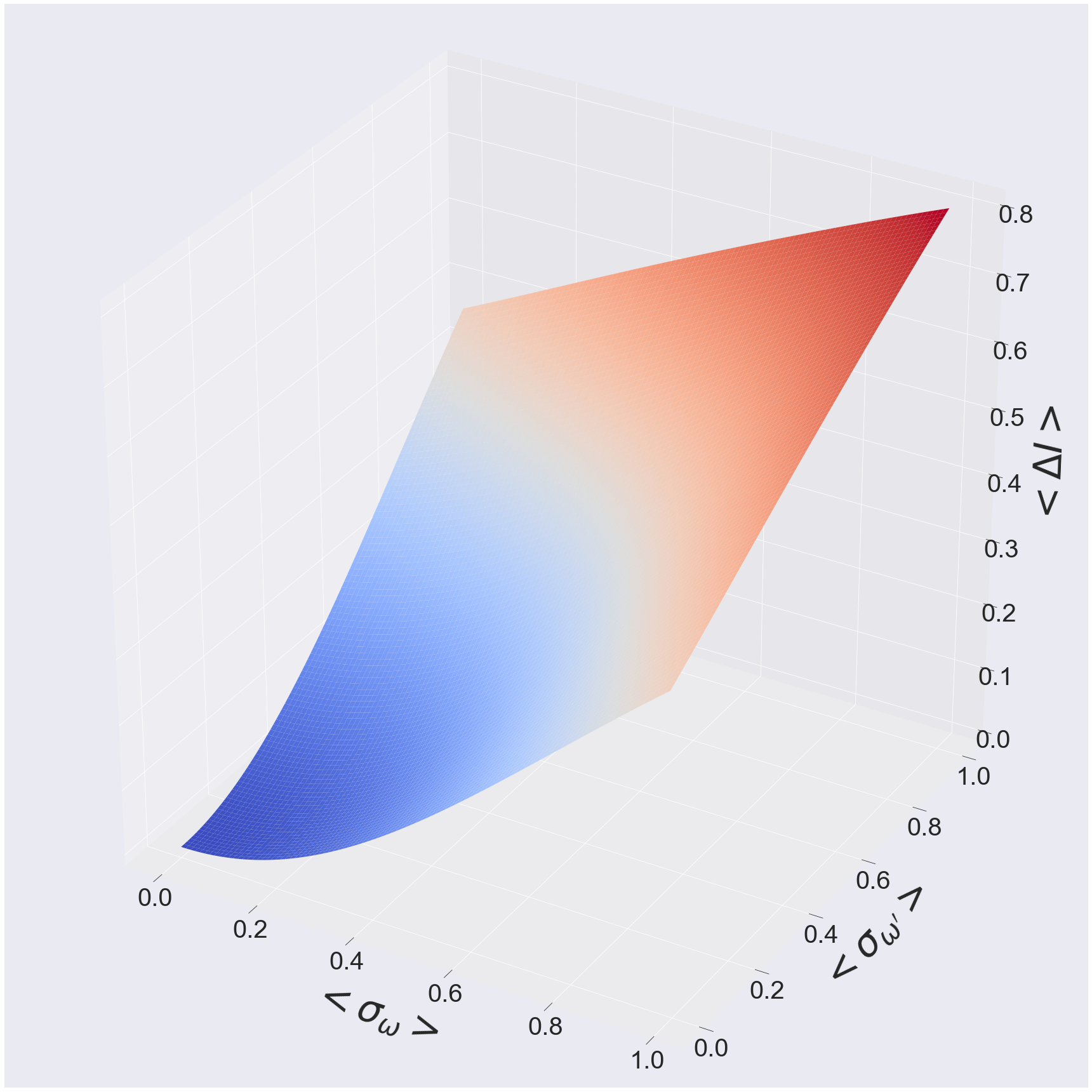}
  \captionof{figure}{We analyze a setting of two non-overlapping units, $\omega$ and $\omega^{\prime}$ that evolve according to a time-symmetric protocol, and the global current precision is fixed. Here, the averaged change in the in-ex information changes to balance the increase (decrease) in the individual averaged EPs over units.}
  \label{fig:test2}
\end{minipage}
\end{figure}

\newpage
\section{DISCUSSION}
\label{sec:conclusion}

In this paper we introduced a way to extend any conventional TUR,
derived for the currents and EP of a single system, to a corresponding 
TUR that instead applies to the case of multiple, dependent systems, by exploiting some of the results in \cite{21, 22, 23}. We illustrated our approach
by deriving the multipartite extensions of the TURs that were derived in \cite{2, 5, 6, 8}. 

There are several avenues of future work suggested by our results. First, it might be possible to use
the same kind of approach to extend the speed-limit theorems \cite{9}, or bounds on first-passage times \cite{24} ---a particular one being dissipation-time uncertainty relation \cite{27}--- to MPPs. 
Note also that the analysis in this paper as well as the analyses in \cite{21, 22, 23} only consider
the case where the same unit structure applies at all times. However, in many real-world scenarios the rate matrices change in way that makes it natural to also choose a unit structure that
changes with time. Integrating such time-dependency of the unit structure into our framework 
may result in tighter bounds on MPPs. 

To conclude, we emphasize that studies on thermodynamic constraints and trade-off relations in MPPs is crucial to obtaining an understanding of the physical systems that are governed by the laws of stochastic thermodynamics. Such systems, including active matter particles that demonstrate aggregation or organelles in a biological cell, are governed by the underlying overall dependency graph combined with the joint dynamics. More work has to be done to elucidate how the topology of a given dependency graph would imply the information-theoretic constraints that are intertwined with the thermodynamic ones. 

\ack
This work was supported in part by the Santa Fe Institute and Grant No. FQXi-RFP-IPW-1912 from the FQXi foundation.
Gülce Kardeş thanks Gianmaria Falasco for helpful discussion.

\appendix
\section{Defining trajectory-level entropy flows}

Trajectory-level EFs must be defined carefully, which
requires that we introduce additional notation that was used in \cite{21, 22, 23}. Write $M(\boldsymbol{x})$ for the total number of state transitions during $\left[0, t_{f}\right]$ made by all subsystems. %\dhwc{Shouldn't that be $\boldsymbol{x}$?}
If $M(\boldsymbol{x}) \geq 1,$ then we define $\eta_{x}:\{1, \ldots, M(\boldsymbol{x})\} \rightarrow \mathcal{N}$ as the function that maps any integer $j \in\{1, \ldots, M(\boldsymbol{x}))\}$ to the subsystem that changes its state by the corresponding transition. We write $k(j)$ for the associated function specifying which reservoir is involved in the $j$'th transition. Similarly, we define $\tau_{x}:\{0, \ldots, M(\boldsymbol{x})\} \rightarrow \mathcal{N}$ to be the function that maps any integer $j \in\{1, \ldots, M(\boldsymbol{x})\}$ to the time of the $j$'th transition (where $0$ is mapped to time $0.$) So $\eta^{-1}(i)$ is the set of all state transitions at which subsystem $i$ changes state in the trajectory $x$. 
(Note that this means that for any set of subsystems $\alpha, \eta^{-1}(\alpha):=\cup_{i \in \alpha} \eta^{-1}(i)$ is the set of all state transitions.) These definitions allow
us to precisely express the \textit{local} entropy flow for any set of subsystems $\alpha$:
\begin{equation}
Q^{\alpha}(\boldsymbol{x})=\sum_{i \in \alpha} \sum_{j \in \eta^{-1}(i)} \ln \left[\frac{W_{\boldsymbol{x}_{r(i)}(\tau(j-1))}^{\boldsymbol{x}_{r(i)}(\tau(j))}(r(i) ; k(j), \tau(j))}{W_{\boldsymbol{x}_{r(i)}(\tau(j))}^{\boldsymbol{x}_{r(i)}(\tau(j-1))}(r(i) ; k(j), \tau(j))}\right].
\end{equation}
\section{Proof of \cref{35}}

MPP FTs, including \cref{34}, were derived in \cite{22}. The joint FT of current and EP, in the form of Eq. (B1), was derived in \cite{5}:
\begin{equation}
\mathbf{P}\left(\sigma,J\right)=e^{\sigma}\tilde{\mathbf{P}}\left(-\sigma,-J\right)
\end{equation}
We can combine these results to obtain joint FTs of vector-valued local currents and vector-valued local EPs in MPPs:
\begin{equation}
\begin{split}
\mathbf{P}\left(\vec{\sigma}^{\mathcal{A}},\vec{J}^{\mathcal{A}}\right) &=\int \mathcal{D} \boldsymbol{x}_{\mathrm{A}} \mathbf{P}\left(\boldsymbol{x}_{\mathrm{A}}\right) \\ & \quad \quad \quad \quad \prod_{\omega \in \mathcal{A}} \delta\left(\sigma^{\omega}-\ln \left[\frac{\mathbf{P}\left(\boldsymbol{x}_{\omega}\right)}{\tilde{\mathbf{P}}\left(\tilde{\boldsymbol{x}}_{\omega}\right)}\right]\right)\\ 
&=e^{\sigma^{\cup \mathcal{A}}} \int \mathcal{D} \boldsymbol{x}_{\mathrm{A}} \tilde{\mathbf{P}}\left(\tilde{\boldsymbol{x}}_{\mathrm{A}}\right) \\
& \quad \quad \quad \quad \prod_{\omega \in \mathcal{A}} \delta\left(\sigma^{\omega}-\ln \left[\frac{\mathbf{P}\left(\boldsymbol{x}_{\omega}\right)}{\tilde{\mathbf{P}}\left(\tilde{\boldsymbol{x}}_{\omega}\right)}\right]\right)\\ 
&=e^{\sigma^{\cup \mathcal{A}}} \int \mathcal{D} \tilde{\boldsymbol{x}}_{\mathrm{A}} \tilde{\mathbf{P}}\left(\tilde{\boldsymbol{x}}_{\mathrm{A}}\right) \\
& \quad \quad \quad \quad \prod_{\omega \in \mathcal{A}} \delta\left(-\sigma^{\omega}-\ln \left[\frac{\tilde{\mathbf{P}}\left(\tilde{\boldsymbol{x}}_{\omega}\right)}{\mathbf{P}\left(\boldsymbol{x}_{\omega}\right)}\right]\right) \\
&=e^{\sigma^{\cup \mathcal{A}}}\tilde{\mathbf{P}}\left(-\vec{\sigma}^{\mathcal{A}},-\vec{J}^{\mathcal{A}}\right).
\end{split}
\end{equation}

\section{Derivation of \cref{38}}

Following \cite{5, 29}, we first define
%We first define, tracing the methodology of \cite{vanvu2019, merhav2010}, 
the distribution $\mathbf{Q}\left(\vec{\sigma}^{\mathcal{A}},\vec{J}^{\mathcal{A}}\right)$:
\begin{equation}
\mathbf{Q}\left(\vec{\sigma}^{\mathcal{A}},\vec{J}^{\mathcal{A}}\right) = \left(1+e^{-\sigma^{\cup \mathcal{A}}}\right)\mathbf{P}\left(\vec{\sigma}^{\mathcal{A}},\vec{J}^{\mathcal{A}}\right)
\end{equation}
This is the joint probability distribution over all components of $\vec{\sigma}^{A}$ and $\vec{J}^{A}$. 
Using this notation, the first moment of a vector-valued current emerging in $\mathcal{A}$ is
\begin{equation}
\mathbf{E}[\vec{J}^{\mathcal{A}}] = \int\displaylimits_{-\infty}^{+\infty}\int\displaylimits_{-\infty}^{+\infty}   \mathbf{P}\left(\vec{\sigma}^{\mathcal{A}},\vec{J}^{\mathcal{A}}\right)\vec{J}^{\mathcal{A}} d\vec{\sigma}^{\mathcal{A}}d\vec{J}^{\mathcal{A}}
\end{equation}

Eq. (C2) captures the individual contributions of currents through each unit in $\mathcal{A}$. Expanding it in terms of $\mathbf{Q}$, we get:
\begin{equation}
\begin{split}
\mathbf{E}[\vec{J}^{\mathcal{A}}] &= \int\displaylimits_{0}^{+\infty}\int\displaylimits_{-\infty}^{+\infty}   \mathbf{Q}\left(\vec{\sigma}^{\mathcal{A}},\vec{J}^{\mathcal{A}}\right)\vec{J}^{\mathcal{A}} \left(\frac{1-e^{-\sigma^{\cup \mathcal{A}}}}{1+e^{-\sigma^{\cup \mathcal{A}}}}\right) d\vec{\sigma}^{\mathcal{A}}d\vec{J}^{\mathcal{A}} \\
&=\left\langle\vec{J}^{\mathcal{A}} \tanh \left(\frac{\sigma^{\cup \mathcal{A}}}{2}\right)\right\rangle_{\mathbf{Q}}
\end{split}
\end{equation}
Accordingly,
\begin{equation}
\mathbf{E}\left[\vec{\sigma}^{\mathcal{A}}\right] =  \left\langle\vec{\sigma}^{\mathcal{A}} \tanh \left(\frac{\sigma^{\cup \mathcal{A}}}{2}\right)\right\rangle_{\mathbf{Q}}
\end{equation}
\begin{equation}
\mathbf{E}\left[\left(\vec{J}^{\mathcal{A}}\right)^2\right] =  \left\langle\left(\vec{J}^{\mathcal{A}}\right)^{2}\right\rangle_{\mathbf{Q}}
\end{equation}
An application of the Cauchy-Schwarz inequality to this result gives
\begin{equation}
\begin{split}
\langle\vec{J}^{\mathcal{A}}\rangle^{2}&=\left\langle\vec{J}^{\mathcal{A}} \tanh \left(\frac{\sigma^{\cup \mathcal{A}}}{2}\right)\right\rangle_{\mathbf{Q}}^{2} 
\\ &\leq\left\langle\left(\vec{J}^{\mathcal{A}}\right)^{2}\right\rangle_{\mathbf{Q}}\left\langle\tanh \left(\frac{\sigma^{\cup \mathcal{A}}}{2}\right)^{2}\right\rangle_{\mathbf{Q}}
\end{split}
\end{equation}

We can now apply the inequalities derived in \cite{5} to Eq.\,(C6):
\begin{equation}
\begin{split}
\left\langle\tanh \left(\frac{\sigma^{ \mathcal{A}}}{2}\right)^{2}\right\rangle_{\mathbf{Q}} &\leq\left\langle\tanh \left[\frac{\sigma^{ \mathcal{A}}}{2} \tanh \left(\frac{\sigma^{ \mathcal{A}}}{2}\right)\right]\right\rangle_{\mathbf{Q}} \\
&\leq \tanh \left(\frac{\left\langle\sigma^{ \mathcal{A}}\right\rangle}{2}\right)
\end{split}
\end{equation}

Combining Eqs.\,(C3)--(C7) we obtain the inequality, $\frac{\left\langle\left(\vec{J}^{\mathcal{A}}\right)^{2}\right\rangle}{\left\langle\vec{J}^{\mathcal{A}}\right\rangle^{2}} \geq \tanh \left(\frac{\langle\sigma^{ \mathcal{A}}\rangle}{2}\right)^{-1}$, which leads to for all units $\alpha_{i}$
\begin{equation}
\frac{\operatorname{Var} J^{\alpha_{i}}}{\left\langle J^{\alpha_{i}}\right\rangle^{2}} \geq \frac{2}{e^{\left\langle\sigma^{A}\right\rangle}-1}.
\end{equation}

\section{Derivation of \cref{37}}

We consider two set of units $\mathcal{A}$ and $\mathcal{B}$. $\vec{J}^{\mathcal{A}}$ indicates the vector whose components are the individual currents ${J}^{\alpha}$ for units $\alpha \in \mathcal{A}$.
Similarly, $\vec{J}^{\mathcal{B}}$ indicates the vector whose components are the individual currents ${J}^{\beta}$ for units $\beta \in \mathcal{B}$. As in Eq.\,(C1), we construct a new distribution from $\mathbf{P}$:
\begin{equation}
\begin{split}
\mathbf{Q}\left(\vec{\sigma}^{\mathcal{A}},\vec{J}^{\mathcal{A}}\right) = \left(1+e^{-\sigma^{\cup \mathcal{A}}}\right)\mathbf{P}\left(\vec{\sigma}^{\mathcal{A}},\vec{J}^{\mathcal{A}}\right) \\
\mathbf{Q}\left(\vec{\sigma}^{\mathcal{B}},\vec{J}^{\mathcal{B}}\right) = \left(1+e^{-\sigma^{\cup \mathcal{B}}}\right)\mathbf{P}\left(\vec{\sigma}^{\mathcal{B}},\vec{J}^{\mathcal{B}}\right)
\end{split}
\end{equation}
We define,
\begin{equation}
\mathbf{E}[\vec{J}^{\mathcal{A}}] =\left\langle\vec{J}^{\mathcal{A}} \tanh \left(\frac{\sigma^{\cup \mathcal{A}}}{2}\right)\right\rangle
\end{equation}
\begin{equation}
\mathbf{E}[\vec{J}^{\mathcal{B}}] =\left\langle\vec{J}^{\mathcal{B}} \tanh \left(\frac{\sigma^{\cup \mathcal{B}}}{2}\right)\right\rangle
\end{equation}

\begin{equation}
\mathbf{E}\left[\vec{\sigma}^{\mathcal{A}}\right] =  \left\langle\vec{\sigma}^{\mathcal{A}} \tanh \left(\frac{\sigma^{\cup \mathcal{A}}}{2}\right)\right\rangle
\end{equation}
\begin{equation}
\mathbf{E}[\vec{J}^{\mathcal{A}}\vec{J}^{\mathcal{B}}] = \int\displaylimits_{-\infty}^{+\infty}\int\displaylimits_{-\infty}^{+\infty}   \mathbf{P}\left(\vec{\sigma},\vec{J}\right)\vec{J}^{\mathcal{A}}\vec{J}^{\mathcal{B}} d\vec{\sigma}d\vec{J}
\end{equation}
Repeating the steps in Appendix C, we write Eq. (D6) as:
\begin{equation}
\mathbf{E}[\vec{J}^{\mathcal{A}}\vec{J}^{\mathcal{B}}] =\left\langle\vec{J}^{\alpha}\vec{J}^{B} \tanh \left(\frac{\sigma^{\cup \mathcal{A}}}{2}\right) \tanh \left(\frac{\sigma^{\cup \mathcal{B}}}{2}\right)\right\rangle
\end{equation}
We note that,
\begin{equation}
\begin{split}
\langle\vec{J}^{\mathcal{A}}\vec{J}^{B}\rangle &\leq\left\langle\vec{J}^{\mathcal{A}}\right\rangle\left\langle\vec{J}^{\mathcal{B}}\right\rangle \times \\ 
&\left\langle\tanh \left(\frac{\sigma^{\cup \mathcal{A}}}{2}\right)\tanh \left(\frac{\sigma^{\cup \mathcal{B}}}{2}\right)\right\rangle
\end{split}
\end{equation}
We then use the product formula for hyperbolic tangents, $\tanh (a+b)=\frac{\tanh (a)+\tanh (b)}{\tanh (a) \tanh (b)+1}$. Hence we express Eq.\,(D8) as follows:
\begin{equation}
\begin{split}
\langle\vec{J}^{\mathcal{A}}\vec{J}^{B}\rangle &\leq\left\langle\vec{J}^{\mathcal{A}}\right\rangle\left\langle\vec{J}^{\mathcal{B}}\right\rangle \times \\
&\left\langle\frac{\tanh \left(\frac{\sigma^{\cup \mathcal{A}}}{2}\right)+\left(\frac{\sigma^{\cup \mathcal{B}}}{2}\right)}{\tanh\left(\frac{\sigma^{\cup \mathcal{A}}+\sigma^{\cup \mathcal{B}}}{2}\right)}-1\right\rangle
\end{split}
\end{equation}
We also note that
\begin{equation}
\begin{split}
&\frac{\tanh \left(\frac{\sigma^{\cup \mathcal{A}}}{2}\right)+\left(\frac{\sigma^{\cup \mathcal{B}}}{2}\right)}{\tanh\left(\frac{\sigma^{\cup \mathcal{A}}+\sigma^{\cup \mathcal{B}}}{2}\right)} - 1 = \\
&1 - \frac{2\cdot\left(e^{-\sigma^{\cup \mathcal{A}}}+e^{-\sigma^{\cup \mathcal{B}}}\right)}{1+e^{-\sigma^{\cup \mathcal{A}}-\sigma^{\cup \mathcal{B}}}+e^{-\sigma^{\cup \mathcal{A}}}+e^{-\sigma^{\cup \mathcal{B}}}}
\end{split}
\end{equation}
Finally, noting that
\begin{equation}
\frac{\langle\vec{J}^{\mathcal{A}}\vec{J}^{B}\rangle-\left\langle\vec{J}^{\mathcal{A}}\right\rangle\left\langle\vec{J}^{\mathcal{B}}\right\rangle}{\left\langle\vec{J}^{\mathcal{A}}\right\rangle\left\langle\vec{J}^{\mathcal{B}}\right\rangle} = \frac{\langle\vec{J}^{\mathcal{A}}\vec{J}^{B}\rangle}{\left\langle\vec{J}^{\mathcal{A}}\right\rangle\left\langle\vec{J}^{\mathcal{B}}\right\rangle}-1
\end{equation}
and combining with the results above, we obtain for any pairs of units $\alpha_{i}$, $\beta_{i}$
\begin{equation}
\begin{split}
&\frac{\operatorname{Covar}\left[J^{\alpha_{i}}, J^{\beta_{j}}\right]}{\left\langle J^{\alpha_{i}}\right\rangle\left\langle J^{\beta_{j}}\right\rangle} \geq \\ 
&\frac{2 \cdot \left(1+e^{\langle\sigma^{\cup \mathcal{A}}-\sigma^{\cup \mathcal{B}}\rangle}\right)}{e^{\langle\sigma^{\cup \mathcal{A}}\rangle}+e^{\langle\sigma^{\cup \mathcal{B}}\rangle}-\left(1+e^{\langle\sigma^{\cup \mathcal{A}}-\sigma^{\cup \mathcal{B}}\rangle}\right)}.
\end{split}
\end{equation}

\end{document}